\documentclass[
a4paper,superscriptaddress,
amsmath,amssymb,10pt,
aps,
floatfix,nofootinbib
]{revtex4-2}

\usepackage{babel}
\usepackage{graphicx}
\usepackage{amsmath,amsfonts,amssymb}

\usepackage{xcolor}

\colorlet{mylinkcolor}{blue!66!black!80}

\definecolor{blind1}{HTML}{C60F7B}
\definecolor{blind2}{HTML}{FFFC31}
\definecolor{blind3}{HTML}{029e73}
\definecolor{blind4}{HTML}{d55e00}
\definecolor{blind5}{HTML}{cc78bc}
\definecolor{blind6}{HTML}{ca9161}
\definecolor{blind7}{HTML}{fbafe4}
\definecolor{blind8}{HTML}{949494}
\definecolor{blind9}{HTML}{ece133}
\definecolor{blind10}{HTML}{56b4e9}
\usepackage{tikz}
\usepackage{gnuplot-lua-tikz}
\usetikzlibrary{calc,arrows,decorations.pathreplacing,backgrounds,shapes,mindmap,fadings}

\usepackage[colorlinks=true,linkcolor=mylinkcolor,citecolor=mylinkcolor,filecolor=cyan,urlcolor=mylinkcolor,breaklinks=true]{hyperref}

\usepackage[utf8]{inputenc}

\usepackage{mathtools}

\renewcommand{\d}{\mathrm{d}}

\newcommand{\avg}[1]{\langle#1\rangle}

\newcommand{\del}{\partial}
\newcommand{\dd}{\mathrm{d}}

\newcommand{\bbb}{\boldsymbol{b}}
\newcommand{\bone}{\boldsymbol{1}}
\newcommand{\bc}{\boldsymbol{c}}

\newcommand{\mA}{\mathbf{A}}
\newcommand{\mB}{\mathbf{B}}

\newcommand{\mD}{\mathbf{D}}
\newcommand{\mI}{\mathbf{I}}

\newcommand{\mL}{\mathbf{L}}

\newcommand{\T}{\top}

\newcommand{\id}{\mathbf{I}}

\newcommand{\twp}{\tilde{\wp}}

\newcommand{\pss}{p^{\rm ss}}

\makeatletter
\def\moverlay{\mathpalette\mov@rlay}
\def\mov@rlay#1#2{\leavevmode\vtop{%
   \baselineskip\z@skip \lineskiplimit-\maxdimen
   \ialign{\hfil$\m@th#1##$\hfil\cr#2\crcr}}}
\newcommand{\charfusion}[3][\mathord]{
    #1{\ifx#1\mathop\vphantom{#2}\fi
        \mathpalette\mov@rlay{#2\cr#3}
      }
    \ifx#1\mathop\expandafter\displaylimits\fi}
\makeatother

\newcommand{\bigcupdot}{\charfusion[\mathop]{\bigcup}{\cdot}}

\newcommand{\dS}{\dot{S}}

\begin{document}
\title[Milestoning Markov processes to thermodynamically consistent renewal dynamics]
{
Milestoning Markov-jump dynamics:
Stationary properties, thermodynamic consistency, kinetic hysteresis,
and fluctuation symmetries}

\author{Tassilo Schwarz}
\affiliation{Mathematical Institute, University of Oxford, OX2 6GG
  Oxford, United Kingdom}
\affiliation{Mathematical bioPhysics Group, Max-Planck-Institute for
  Multidisciplinary Sciences, G\"{o}ttingen 37077, Germany}

\author{David Hartich}
\affiliation{Mathematical bioPhysics Group, Max-Planck-Institute for
  Multidisciplinary Sciences, G\"{o}ttingen 37077, Germany}

\author{Alja\v{z} Godec}
\email{agodec@physik.uni-freiburg.de}
\affiliation{Mathematical Physics and Stochastic Dynamics, Institute
  of Physics, University of Freiburg, 79104 Freiburg im Breisgau,
  Germany}
\affiliation{Mathematical bioPhysics Group, Max-Planck-Institute for
  Multidisciplinary Sciences, G\"{o}ttingen 37077, Germany}

\begin{abstract}
We derive an exact coarse graining of generic Markov-jump processes into observable semi-Markov dynamics. Exact results for waiting-time distributions for jumps between observable states are derived and proved that these decompose into conditionally independent dwell and transition times.  Dwell times are proved to be a local property of mesostates --- they depend on the initial but not final state. Conversely, transition-path times depend on both states, trigger kinetic hysteresis, and, under suitable conditions on the hidden sub-network, are shown to obey a reflection symmetry. We characterize the stationary properties of the milestoned dynamics, prove its thermodynamic consistency, and demonstrate robustness to milestone positioning. Surprisingly, even in the limit of a time-scale separation rendering the observed dynamics approximately Markovian, the effect of kinetic hysteresis on the dissipation persists.  A minimal example shows how the results lay the foundation for inferring affinities of hidden dissipative cycles from observations of transition-path times.
\end{abstract}
 
\maketitle

\section{Introduction}

The dynamics of many biophysical systems on relevant time scales, 
such as gene-expression \cite{paul05,mcad97}, chemical reactions \cite{gill07,gill77,elbe20,polettiniDissipationNoisyChemical2015,raoNonequilibriumThermodynamicsChemical2016}, molecular machines \cite{seif12}, protein dynamics \cite{chod14,bowm10,wood14}, or neuroscience \cite{lynnBrokenDetailedBalance2021a}, is theoretically well described by continuous-time Markov dynamics. 
Typically, however, one can experimentally observe only coarse-grained versions 
of the dynamics that generally display memory \cite{Zwanzig1973JSP,Mori1965PTP,Lapolla2019FP,zhaoEmergenceMemoryEquilibrium2024}, because measurements on complex systems either have a finite resolution or they simply cannot resolve all degrees of freedom \cite{dieballPerspectiveTimeIrreversibility2025}. 
Such non-Markovian mesoscopic observations raise important challenges, in particular in the context of thermodynamics \cite{schwarzConsistentTimeReversal2025,PhysRevX.12.041026,dieballThermodynamicBoundsGeneralized2024,dieballPerspectiveTimeIrreversibility2025,stutzerStochasticCalculusPathwise2025,zhaoEmergenceMemoryEquilibrium2024, blomMilestoningEstimatorsDissipation2024a, hartichEmergentMemoryKinetic2021a,hartichViolationLocalDetailed2023,seifertUniversalBoundsEntropy2025,fritzEntropyEstimationPartially2025,maierInferringKineticsEntropy2024} (for a recent account see \cite{dieballPerspectiveTimeIrreversibility2025}).

In the case of 
dynamics evolving on a graph the two dominant forms of coarse graining are lumping, where only a mesoscopic partition of the microscopic state space is observable \cite{Esposito2012PRE,Seifert_2019,skin21,PhysRevLett.127.198101,Hartich2023PRR,dieballPerspectiveTimeIrreversibility2025,tabanera-bravoPurelyQuantumMemory2025a,zhaoMarkovstateHolography2025,schwarzConsistentTimeReversal2025,raha07},
and milestoning, where only a subset of the microscopic states is observable \cite{hartichEmergentMemoryKinetic2021a,Hartich2023PRR,blomMilestoningEstimatorsDissipation2024a,berezhkovskiiNonequilibriumSingleMoleculeTrajectories2020,vandermeerThermodynamicInferencePartially2022a}.
Even if the underlying microscopic dynamics are indeed Markovian, the coarse-graining leads to non-Markovian observations \cite{maierCompensatingRandomTransitiondetection2025, vandermeerThermodynamicInferencePartially2022a, schwarzConsistentTimeReversal2025,mart19a,hartichCommentInferringBroken2024}. In particular, when milestoning a Markov into a semi-Markov process, the sequence of visited mesostates is Markovian whereas the waiting times in the mesostates cease to be exponentially distributed and depend on the next-visited state \cite{hartichEmergentMemoryKinetic2021a}. Note that Markov dynamics are renewal dynamics, but the converse is untrue. Dynamical fluctuations in the form of an Onsager-Machlup theory of semi-Markov processes were developed in \cite{Maes_2009}, and 
the distinction between violations of detailed balance and
the violation of a ``mathematical'' time-reversal symmetry in
semi-Markov processes was highlighted in \cite{wang07} and deepened in \cite{hartichCommentInferringBroken2024}.  
Milestoning general diffusion processes on a graph was shown to lead (in stark contrast to lumping) to a thermodynamically consistent semi-Markov dynamics with a consistent Markovian limit under a time-scale separation \cite{hartichEmergentMemoryKinetic2021a,Hartich2023PRR}. Corresponding general results for milestoning Markov-jump dynamics remained elusive.

In this work, we formulate a thermodynamically consistent renewal theory for coarse-grained dynamics emerging from milestoning of continuous-time Markov jump processes, in analogy to the continuous-space setting \cite{hartichEmergentMemoryKinetic2021a}, and generalize it to a more complex topology. The resulting semi-Markovian kinetics are based on first-passage time conditioned on the succeeding milestone. We prove that the dwell times in the milestones are local and that the transition path times between any pair of adjacent milestones are symmetric in law. We dissect transitions away from an observed state, whereby we explain the emergence of non-Markovianity.
We show that kinetic hysteresis 
occurs for non-trivial transition times, leading to coarse-graining ceasing to commute with time-reversal. We further show that kinetic hysteresis persists even under time-scale separation.
Under the condition of local detailed balance for each microscopic path between any pairs of milestones, we prove thermodynamic consistency and extend our theory to parallel transition pathways. With explicit examples of dynamics on graphs, we illustrate our analytic results.

\section*{Results}

\subsection{Setup}\label{sec:model}
We assume that the microscopic dynamics corresponds to an irreducible ergodic Markov-jump process in continuous time. The probability $P_x(t)$ to find the system  in the state  $x\in \Omega$ evolves with the master equation
\begin{equation}
 \del_tP_x(t)=\sum_{x'\in \Omega}\big[P_{x'}(t)w_{x'\to x}-P_{x}(t)w_{x\to x'}\big]
 \equiv \sum_{x'\in \Omega} L_{xx'}P_{x'}(t),
 \label{eq:full_master_equation}
\end{equation}
where in the last step we defined the generator $\mL$ with off-diagonal elements $L_{xx'}=w_{x'\to x}$ for $x\neq x'$ and with diagonal $L_{xx}=-\sum_{x'\neq x}w_{x\to x'}$. Note that the probability is preserved, since the vector $\bone\equiv (1,\ldots,1)$ is the left eigenvector $\bone \,\mL=0$. We denote the $i$-th standard basis vector by $\boldsymbol{e_i}$. Throughout this work, we assume \emph{thermodynamic consistency of the microscopic process}, i.e.\ a transition $w_{x' \rightarrow x}>0$ happens with positive probability if and only if the reverse is possible $w_{x \rightarrow x'}>0$.

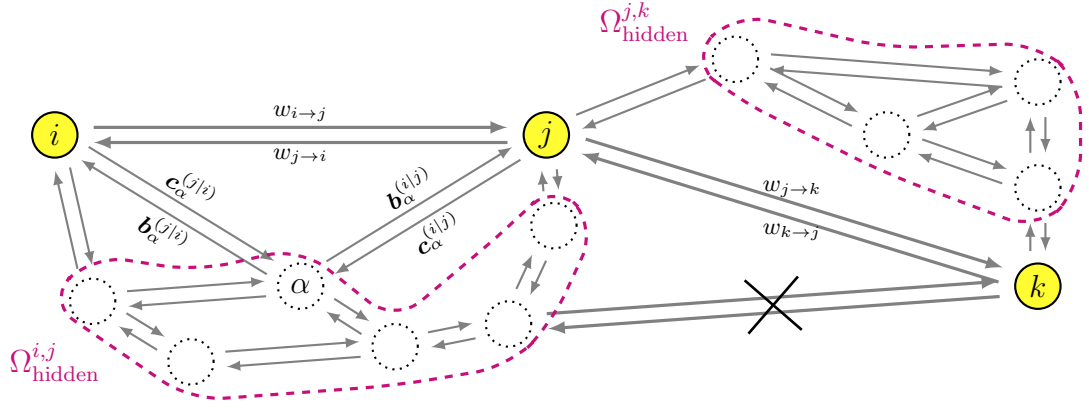
\begin{figure}
\begin{tikzpicture}

\pgfdeclaredecoration{sl}{initial}{
  \state{initial}[width=\pgfdecoratedpathlength-1sp]{
     \pgfmoveto{\pgfpointorigin}
  }
  \state{final}{
     \pgflineto{\pgfpointorigin}
    }
}

\tikzset{parallel arrow/.style={-latex,thick, 
     shorten >=1.5mm, shorten <=1.5mm, 
     decoration={sl,raise=1mm},decorate}}

\tikzset{thick parallel arrow/.style={-latex,very thick, 
     shorten >=2mm, shorten <=2mm, 
     decoration={sl,raise=1mm},decorate}}
\tikzset{
  crossed pair/.style={
    black,
    line width=1pt
  }
}

\newcommand{\crossedge}[2]{%
  \coordinate (crossmid) at ($(#1)!0.5!(#2)$);
  \draw[crossed pair]
    ($(crossmid)!5mm!45:(#2)$)  -- ($(crossmid)!5mm!225:(#2)$)
    ($(crossmid)!5mm!-45:(#2)$) -- ($(crossmid)!5mm!135:(#2)$);
}

	 \node[draw,thick,circle,fill=blind2,minimum size =0.6cm] at  (0,0) (i) {};
	 \node[draw,thick,circle,fill=blind2,minimum size =0.6cm] at  (6.5,0) (j) {};

  \node[draw,thick,circle,fill=blind2,minimum size =0.6cm] at  (13,-2) (k) {};
	 
	  \node[draw,thick, dotted,circle,minimum size =0.6cm] at  (3.25,-2) (a) {};
	  \node[draw,thick, dotted,circle,minimum size =0.6cm] at  (0.5,-2.2) (u1) {};
	  \node[draw,thick, dotted,circle,minimum size =0.6cm] at  (1.8,-3.0) (u2) {};
	  
	 \node[draw,thick, dotted,circle,minimum size =0.6cm] at  (4.5,-2.8) (u3) {};
	 \node[draw,thick, dotted,circle,minimum size =0.6cm] at  (6,-2.5) (u4) {};
	 \node[draw,thick, dotted,circle,minimum size =0.6cm] at  (6.6,-1.2) (u5) {};

    \node[draw,thick, dotted,circle,minimum size =0.6cm] at  (9,1) (o1) {};
    \node[draw,thick, dotted,circle,minimum size =0.6cm] at  (11,0) (o2) {};\node[draw,thick, dotted,circle,minimum size =0.6cm] at  (13,0.7) (o3) {};
    \node[draw,thick, dotted,circle,minimum size =0.6cm] at  (13,-0.7) (o4) {};

	 \path[black!50] (i) edge[thick parallel arrow] node[above=0.5mm,black]{$w_{i\to j}$} (j);
	 \path[black!50] (j) edge[thick parallel arrow] node[below=0.5mm,black]{$w_{j\to i}$} (i);
	 
  \path[black!50] (j) edge[thick parallel arrow] node[above=0.5mm,black]{$w_{j\to k}$} (k);
	 \path[black!50] (k) edge[thick parallel arrow] node[below=0.5mm,black]{$w_{k\to j}$} (j);

\path[black!50] (u4) edge[thick parallel arrow] (k);
\path[black!50] (k)  edge[thick parallel arrow] (u4);
\crossedge{u4}{k}
	 
	 \begin{scope}[black!50]
	 \path (i) edge[parallel arrow] node[above=0.5mm,sloped,black]{$\boldsymbol{c}_{\alpha}^{(j|i)}$} (a);
	 \path (a) edge[parallel arrow] node[below=0.5mm,sloped,black]{$\boldsymbol{b}_{\alpha}^{(j|i)}$} (i);
	 \path (j) edge[parallel arrow] node[above=0.5mm,sloped,black]{$\boldsymbol{b}_{\alpha}^{(i|j)}$} (a);
	 \path (a) edge[parallel arrow] node[below=0.5mm,sloped,black]{$\boldsymbol{c}_{\alpha}^{(i|j)}$} (j);
	 
	 \path (i) edge[parallel arrow]  (u1);
	 \path (u1) edge[parallel arrow]  (i);
	 
	 \path (a) edge[parallel arrow]  (u1);
	 \path (u1) edge[parallel arrow]  (a);
	 
	 \path (u1) edge[parallel arrow]  (u2);
	 \path (u2) edge[parallel arrow]  (u1);
	 
	 \path (u2) edge[parallel arrow]  (u3);
	 \path (u3) edge[parallel arrow]  (u2);
	 
	 \path (a) edge[parallel arrow]  (u3);
	 \path (u3) edge[parallel arrow]  (a);
	 
	 \path (u3) edge[parallel arrow]  (u4);
	 \path (u4) edge[parallel arrow]  (u3);
	 
	 \path (u4) edge[parallel arrow]  (u5);
	 \path (u5) edge[parallel arrow]  (u4);

      \path (o1) edge[parallel arrow]  (o2);
	   \path (o2) edge[parallel arrow]  (o1);

      \path (o1) edge[parallel arrow]  (o3);
	   \path (o3) edge[parallel arrow]  (o1);

      \path (o2) edge[parallel arrow]  (o3);
	   \path (o3) edge[parallel arrow]  (o2);

      \path (o3) edge[parallel arrow]  (o4);
	   \path (o4) edge[parallel arrow]  (o3);

      \path (o2) edge[parallel arrow]  (o4);
	   \path (o4) edge[parallel arrow]  (o2);

  \path (j) edge[parallel arrow]  (u5);
	 \path (u5) edge[parallel arrow]  (j);

      \path (j) edge[parallel arrow]  (o1);
	   \path (o1) edge[parallel arrow]  (j);

     \path (k) edge[parallel arrow]  (o4);
	   \path (o4) edge[parallel arrow]  (k);
	 
	 \draw[blind1,very thick, dashed, rounded corners =4.5mm] ($(u1.west)-(0.3,0)$) --($(u1.north)+(0,0.3)$) --($0.5*(u1)+0.5*(a)+(0,0.3)$)--  ($(a.north)+(0,0.2)$)--(u3.north)--($(u5.west)-(0.26,0)$)--($(u5.north)+(0,0.26)$)--($(u5.east)+(0.25,0)$) -- ($(u4.east)+(0.25,0)$)--($(u4.south)-(0,0.25)$)--($(u3.south)-(0,0.20)$)--($(u2.south)-(0,0.20)$)--cycle;
  \draw[blind1,very thick, dashed, rounded corners =4.5mm] ($(o1.west)-(0.3,0)$) --($(o1.north)+(0,0.3)$) --($0.5*(o1)+0.5*(o3)+(0,0.3)$)--  ($(o3.north)+(0,0.2)$)--($(o3.east)+(0.25,0)$) -- ($(o4.east)+(0.25,0)$)--($(o4.south)-(0,0.25)$)--($(o2.south)-(0,0.20)$)--($(o1.south)-(0,0.20)$)--cycle;

  \node[blind1,font=\large] at (0,-3) {$\Omega_{\rm hidden}^{i,j}$};
  \node[blind1,font=\large] at (7.8,1.5) {$\Omega_{\rm hidden}^{j,k}$};
	 
	 \end{scope}
	\node at (i) {\large $i$};
	\node at (j) {\large $j$}; 
	\node at (k) {\large $k$}; 
	
	\node at (a) {\large $\alpha$};
\end{tikzpicture}
 \caption{Definition of the model. Network with two observed states $i,j$. The decomposition into observed nodes and hidden ones. Any pair of observed nodes $i$ and $j$ from $\Omega_{\rm obs}$ are connected directly via direct transition rates ($w_{i\to j}$ and $w_{j\to i}$) and/or via a set of hidden nodes $\Omega_{\rm hidden}^{i,j}=\Omega_{\rm hidden}^{i,j}$. We assume that all hidden states $\alpha\in\Omega_{\rm hidden}^{i,j}$ are \emph{not directly} connected to a third observed state $k\neq i,j$ (crossed out edges).}
 
 \label{fig:cg_subnetworks}
\end{figure}

We now partition the full set of network states $\Omega$ into a set of observed states, $\Omega_{\rm obs}$, and a set of unobserved hidden states, $\Omega_{\rm hidden}$.
In milestoning the observed states are called ``cores'' \cite{schu11,bere19} (see also \cite{fara04,shal06,elbe20,elbe20a,suar21}).
The hidden states are assumed to connect  a pair of observed states in the following manner: Starting from any hidden node $\alpha\in \Omega_{\rm hidden}$ we assume one can reach exactly two nodes $i,j\in \Omega_{\rm obs}$, without passing by another observed node in between (note the forbidden transition in Fig.~\ref{fig:cg_subnetworks}). From now on we use for the observed nodes the symbols $i,j,k\in \Omega_{\rm obs}$, whereas Greek letters $\alpha,\beta, \gamma \in \Omega_{\rm hidden}$ indicate hidden states.

The set of hidden states is represented by $\Omega_{\rm hidden}=\bigcup_{i,j}\Omega_{\rm hidden}^{i,j}$, where
 $\Omega_{\rm hidden}^{i,j}=\Omega_{\rm hidden}^{j,i}$ is the set of hidden network states that connects a pair of observed nodes $i$ and $j$. 
 Note that $\Omega_{\rm hidden}^{i,j}$ may well be an empty set $\Omega_{\rm hidden}^{i,j}=\emptyset$ if $i$ and $j$ are connected solely via direct transitions ($w_{i\to j},w_{j\to i}\neq0$) or if $i$ and $j$ are not connected at all.
If $\Omega_{\rm hidden}^{i,j}$ is nonempty, there exists at least one element $\alpha\in\Omega_{\rm hidden}^{i,j}$
which is connected to $i$, that is, $w_{i\to\alpha }\neq0$ and $w_{\alpha\to i }\neq0$. 
Moreover, the rates along all paths $(i=x_0\to x_1\to\ldots\to x_n=j)$ on the subset
$\Omega_{\rm hidden}^{i,j}\cup\{i,j\}$ from state $x_0=i$ to state $x_n=j$ $(n\ge1)$
are assumed to satisfy
\begin{equation}
\sum_{\tau=0}^{n-1}\ln\frac{w_{x_{\tau}\to x_{\tau+1}}}{w_{x_{\tau+1}\to x_{\tau}}}=\mathcal{A}_{ij} 
\label{eq:quasiLDB}
\end{equation} 
where $\mathcal{A}_{ij}$ is the affinity from observed state $i$ to observed state $j$. If the full master equation \eqref{eq:full_master_equation} satisfies detailed balance, one can find a free energy function $E_i$ satisfying $\mathcal{A}_{ij}=E_i-E_j$ for all $i,j\in \Omega_{\rm obs}$. In Sec.~\ref{sec:generalization} we drop the restriction of Eq.~\eqref{eq:quasiLDB} by allowing multiple mechanisms $\nu$ to transition from $i$ to $j$, so that we will then shift from $\mathcal{A}_{ij}$ to  $\mathcal{A}_{ij}(\nu)$.

The observed dynamics away from the observed node $i$ becomes \emph{locally}
non-Markovian as soon as there are adjacent hidden states, that is, when $\bigcup_j\Omega_{\rm hidden}^{i,j}\neq \emptyset$. The goal of this article  will be to investigate the local dynamics on the subset $\Omega^{(i)}\equiv(\bigcup_j\Omega_{\rm hidden}^{i,j})\cup\{i\}$ for each observed state $i\in \Omega_{\rm obs}$ to fully quantify the non-Markovian  dynamics between all pairs of observed states ($i \to j$).

\subsection{Coarse graining Markov to semi-Markov process -- Milestoning}
\subsubsection{Definition of local kinetics away from an observed state}
In this subsection we  specify the non-Markovian waiting time distribution away from an observed node
$i$ to another observed state $j$. The dynamics has three contributions
\begin{itemize}
 \item[ \bfseries(i)] intrinsic jumps within each set of hidden states $\Omega_{\rm hidden}^{i,j}$,
 \item[ \bfseries(ii)] jumps between the observed and hidden states ``$\{i,j\}\to\Omega_{\rm hidden}^{i,j}$'',
 \item[ \bfseries(iii)] direct jumps between pairs of observed states ($i\to j$ or $j\to i$).
\end{itemize}
Note that the case of Markovian dynamics without hidden states reduces to having only contributions of type (iii).
The sequential approach we follow here is key to understanding the emergence of non-Markovian network dynamics.\footnote{The second and last key step will be the identification of transition paths.}  To that end we analyze the kinetics on the subset $\Omega^{(i)}$ and address items (i)-(iii) paragraph-by-paragraph.

Let us begin with item (i), the intrinsic jumps $\alpha\to\beta$ within  $\alpha,\beta\in\Omega_{\rm hidden}^{i,j}$ which occur at rate $w_{\alpha\to\beta}$.
Therefore, we consider the generator of the dynamics on the induced subgraph on $\Omega_{\rm hidden}^{i,j}$ (see dotted pink region in Fig.~\ref{fig:cg_subnetworks}). That is, the generator $\mL^{i,j}$ describing dynamics only along  hidden transitions  within $\Omega_{\rm hidden}^{i,j}$ defined for $\alpha,\beta\in \Omega_{\rm hidden}^{i,j}$ by
\begin{equation}
 L^ {i,j}_{\beta \alpha}\equiv
 \begin{cases}
 w_{\alpha\to\beta}, &
 \alpha\neq\beta\\
 -\!\!\!\!\!\displaystyle{\sum_{\gamma\in \Omega^{i,j}_{\rm hidden}\backslash\{\alpha\}}\!\!\!\!\!\!\!\!\!\!\!\!w_{\alpha\to\gamma}}, & \alpha=\beta.
\end{cases} 
\label{eq:L_hidden}
\end{equation}
Due to $\Omega_{\rm hidden}^{i,j}=\Omega_{\rm hidden}^{j,i}$, the induced subgraphs are equivalent, and hence this generator satisfies $\mL^{j,i}=\mL^{i,j}$. 
The generator of eq.~\eqref{eq:L_hidden} preserves
probability,
$\bone\,\mL^{i,j}=0 $. Condition~\eqref{eq:quasiLDB} implies that the generator alone $\mL^{i,j}$ satisfies detailed balance \cite{kamp07}, that is,
there exists an invariant state $\boldsymbol{\pi }$ satisfying
$L^{j,i}_{\beta\alpha}\pi_\alpha=L^{j,i}_{\alpha\beta}\pi_\beta$ for all $\alpha, \beta \in \Omega_{\rm hidden}^{i,j}$.

Let us now turn to item (ii), the jumps between observed and hidden nodes.
That is, transitions between $i,j$ and $\alpha\in\Omega_{\rm hidden}^{i,j}$.
Let us first focus on transitions from hidden back to observed nodes,
which we denote through the vectors $\bbb^{(j|i)}$,  $\bbb^{(i|j)}$ (see Fig.~\ref{fig:cg_subnetworks}) with components
\begin{equation}
 \bbb^{(j|i)}_\alpha=w_{\alpha\to i}\qquad\text{and} \qquad \bbb^{(i|j)}_\alpha=w_{\alpha\to j}\qquad\forall \alpha \in\Omega_{\rm hidden}^{i,j}.
  \label{eq:hidden>obs_vec}
\end{equation}
Similarly, the transitions from observed to hidden nodes can be combined into a vector $\bc^{(j|i)}$ with elements
\begin{equation}
 \bc^{(j|i)}_\alpha=w_{i\to \alpha}\qquad\text{and} \qquad \bc^{(i|j)}_\alpha=w_{j\to \alpha}\qquad\forall \alpha \in\Omega_{\rm hidden}^{i,j}.
 \label{eq:obs>hidden_vec}
\end{equation}
It will prove convenient to introduce the diagonal matrices
\begin{equation}
 \mathbf{B}^{(j|i)}=\operatorname{diag}[\bbb^{(j|i)}]\qquad  \text{and}\qquad  \mathbf{C}^{(j|i)}=\operatorname{diag}[\bc^{(j|i)}]
 \label{eq:hidden<>obs_mat}
\end{equation}
which satisfy $\bbb^{(j|i)}= \mathbf{B}^{(j|i)}~\bone^\T$ and $\bc^{(j|i)}= \mathbf{C}^{(j|i)}~\bone^\T$.

The last contribution according to item (iii) stems from the direct transitions. All direct transitions
away from the observed node $i$ are occurring with the total rate 
\begin{equation}
 r^{(i)}=\sum_{j\in \Omega_{\rm obs}\backslash\{i\}}w_{i\to j}.
\end{equation}
This allows us to write the generator $\mL^{(i)}$ on the states  $\Omega^{(i)} \coloneq \{i\}\cup(\bigcup_j\Omega_{\rm hidden}^{i,j})$, i.e.\ on the state $i$ together with all hidden states which are visited on any path from $i$ before reaching any other visible state, 
in the following block matrix structure
\begin{align}
 \mL^{(i)}
 &=\mL^{(i)}_{0}-\boldsymbol{\mathcal{D}}^{(i)}\nonumber\\
 &\equiv
 \begin{pmatrix}
   -c^{(i)}&(\bbb^{(j_1|i)})^\T&\cdots& (\bbb^{(j_N|i)} )^\T\\
 \bc^{(j_1|i)}&\mL^{j_1,i}-\mB^{(j_1|i)}\\
  \vdots&&\ddots\\
   \bc^{(j_N|i)}&&&\mL^{j_N,i}-\mB^{(j_N|i)}
 \end{pmatrix}
-\begin{pmatrix}
   r^{(i)}&&\\
&\mB^{(i|j_1)}\\
&&\ddots\\
&&& \mB^{(i|j_N)}
 \end{pmatrix}
 ,
 \label{eq:generator}
\end{align} where $j_1, \ldots, j_N$ is an enumeration of the observable vertices reachable from $i$ through only hidden states, i.e.\ of the  set $\bigcup_j\Omega_{\rm hidden}^{i,j}$.
Each block matrix is defined as follows. The first column in the first block matrix, $\mL^{(i)}_{0}$,
in \eqref{eq:generator} contains the jumps from $i$ to the hidden nodes with $\bc^{(j|i)}$
from   Eq.~\eqref{eq:obs>hidden_vec}, while $c^{(i)}$ is their sum $c^{(i)}=\sum_{j\neq i} \bone \,\bc^{(j|i)}\equiv\sum_{j\neq i} \|\bc^{(j|i)}\|_1$. The first row in the first block matrix
in \eqref{eq:generator} contains the jumps from the hidden nodes back to the observed node $i$
as defined in \eqref{eq:hidden>obs_vec}, while the remainder is a block diagonal matrix
with diagonal blocks ``$\mL^{i,j}-\mB^{(i|j)}$'', where the generator $\mL^{i,j}$
is defined in Eq.~\eqref{eq:L_hidden} and $\mB^{(i|j)}$ is defined in Eqs.~\eqref{eq:hidden>obs_vec} and \eqref{eq:hidden<>obs_mat}. Note that the
 first block matrix
in \eqref{eq:generator} preserves the probability on the subset $\Omega^{(i)}$ since
$\bone \,\mL^{(i)}_{0}=0$
due to $\bone \,\mB^{(j|i)}=(\bbb^{(j|i)})^\T$ [cf. Eqs.~\eqref{eq:hidden>obs_vec} and \eqref{eq:hidden<>obs_mat}]. The second block  matrix $\boldsymbol{\mathcal{D}}^{(i)}$
is a diagonal matrix solely containing transition rates exiting the subset $\Omega^{(i)}$. In the following we exploit the block matrix structure in Eq.~\eqref{eq:generator} to efficiently analyze the non-Markovian kinetics between \emph{observed}  state changes.

\subsubsection{Waiting time distribution between an observed state change (coarse-graining)}
The waiting time distribution between observed state change away from an observed node $i$ can be characterized by the survival probability that starting from $i$ another observed node has not yet been visited. The survival probability is given by
\begin{equation}
 \mathcal{P}_i(t)=
\bone\exp({\mL^{(i)}t})\boldsymbol{e}_1,
\label{eq:surv_def}
\end{equation}
where $\mL^{(i)}$ is the generator defined in Eq.~\eqref{eq:generator} and
$\boldsymbol{e}_1=(1,0,\ldots,0)^\T$ is the first vector of the standard basis.

The survival probability decays monotonically from $\mathcal{P}_i(0)=1$
to $\mathcal{P}_i(t)\to 0$ in the limit $t\to\infty$. The probability \emph{density} of the waiting time (to exit state $i$) is the rate at which the survival probability decays
\begin{equation}
\wp^{\rm exit}_i(t)\equiv-\del_t\mathcal{P}_i(t)=-\bone \mL^{(i)}\exp(\mL^{(i)}t) \boldsymbol{e}_1
 =
  \bone \boldsymbol{\mathcal{D}}^{(i)}\exp(\mL^{(i)}t) \boldsymbol{e}_1,
 \label{eq:exit_pre}
\end{equation}
where in the second step we used Eq.~\eqref{eq:surv_def}, and in the last step we used
Eq.~\eqref{eq:generator} and $\bone \,\mL^{(i)}_0= 0$.
The decomposition of generator \eqref{eq:generator} allows us to rewrite the
density of the exit time in the form
\begin{equation}
 \wp^{\rm exit}_i(t)=\bone \boldsymbol{\mathcal{D}}^{(i)}\exp({\mL^{(i)}t})\boldsymbol{e}_1=
 \begin{pmatrix}
  r^{(i)}\\\bbb^{(i|1)}\\\vdots\\ \bbb^{(i|N)}
 \end{pmatrix}^\T
\exp({\mL^{(i)}t})
 \begin{pmatrix}
  1\\0\\\vdots\\0
 \end{pmatrix}
,
 \label{eq:exit}
\end{equation}
where in the second step we inserted \eqref{eq:generator} along with $\bbb^{(i|j)}=\mB^{(i|j)}~\bone^\T$. Note that the probability density is normalized  $\int_0^\infty\wp^{\rm exit}_i(t)\dd t=1$.

Eq.~\eqref{eq:exit_pre} only characterizes the statistics of leaving state $i$. A change of state corresponds to leaving state $i$ \emph{and} entering a new state $j$ at time $t$. The joint density will be denoted by $\wp_{j|i}(t)=\text{Prob}(t,j|i)$ and is normalized according to $\sum_{j \neq i}\int_0^\infty\wp_{j|i}(t)\dd t$. The exit time density follows from its marginalization $ \wp^{\rm exit}_i(t)=\sum_{j\neq i}\wp_{j|i}(t)$.
Inspecting Eq.~\eqref{eq:exit} and using $r^{(i)}=\sum_{j \neq i}w_{i\to j}$ we find that the joint density\footnote{Joint in the sense of jointly between waiting time and splitting.} is given by
\begin{equation}
 \wp_{j|i}(t)=
\left(
 w_{i\to j},0,\dots,0,(\bbb^{(i|j)})^\T,0,\ldots,0\right)
\exp(\mL^{(i)}t)
 \boldsymbol{e}_1
,
 \label{eq:wp}
\end{equation}
where $\bbb^{(i|j)}$ contains all transitions to node $j$ from hidden ``intermediate'' states from $\Omega_{\rm hidden}^{i,j}$. 
It proves convenient to Laplace transform\footnote{For a generic function $f(t)$ the Laplace transform is defined by $\tilde f(s)=\int_0^\infty\exp(-st)f(t)\dd t$.},
which renders Eq.~\eqref{eq:wp}
into

\begin{align}
 \tilde{\wp}_{j|i}(s)= \mathcal L\left[\wp_{j|i}\right](s)=
\left(
 w_{i\to j},0,\dots,0,(\bbb^{(i|j)})^\T,0,\ldots,0\right)
\Big[\id s-\mL^{(i)}\Big]^{-1}
 \boldsymbol{e}_1,
 \label{eq:lp-of-cond-trans-dens}
\end{align}
where $\id$ is the identity matrix.

The inverse of a generic block matrix $\mathbf{\mathcal{R}}$ with invertible $\mD$ and invertible Schur complement $m:=a-\bbb^\T \mD^{-1}\bc$
is given by \cite{lu02} (where the right blocks are irrelevant since we multiply by $\boldsymbol{e}_1$)
\begin{equation}
\mathbf{\mathcal{R}}^{-1}\equiv
\begin{pmatrix}
 a&-\bbb^\T\\
 -\bc&\mD
\end{pmatrix}^{-1}
=
\begin{pmatrix}
 \frac{1}{m}& *\\
  \frac{1}{m}\mD^{-1}\bc&*
\end{pmatrix}
\end{equation}
with coefficient $a$, vectors $\bbb,\bc$ and square matrix $\mD$. Identifying ``$\mathbf{\mathcal{R}}=\id s-\mL^{(i)}$''
we obtain, by identifying for example $a=s+r^{(i)}+c^{(i)}$,
\begin{align}
\twp_{j|i}(s)
&=\frac{w_{i \to j} + (\bbb^{(i \mid j)})^\T \mD^{-1} \bc }{m} \\
&=\frac{ w_{i\to j}+(\bbb^{(i|j)})^\T\big[\id s+\mB^{(j|i)}+\mB^{(i|j)}-\mL^{i,j}\big]^{-1}\bc^{(j|i)}}{s+r^{(i)} +c^{(i)}-\sum_k(\bbb^{(k|i)})^\T\big[\id s+\mB^{(k|i)}+\mB^{(i|k)}-\mL^{k,i}\big]^{-1}\bc^{(k|i)}}.
 \label{eq:twp_pre}
\end{align}
Note in the numerator the term $\bbb^{(i|j)}$ has the initial node $i$ as first component, whereas
  the denominator has $\bbb^{(k|i)}$. Using the identities $(\bbb^{(k|i)})^\T=\bone \,\mB^{(k|i)}$, $\bone \,\mL^{k,i}=0$, and $c^{(i)}=\sum_k \bone \,\bc^{(k|i)}$,
 Eq.~\eqref{eq:twp_pre} after some tedious but straightforward calculation becomes
\begin{align}
\twp_{j|i}(s)
=
\frac{
w_{i\to j}+(\bbb^{(i|j)})^\T\big[\id s+\mB^{(j|i)}+\mB^{(i|j)}-\mL^{i,j}\big]^{-1}\bc^{(j|i)}
}{
s+r^{(i)} +\sum_k\big[s\bone+(\bbb^{(i|k)})^\T\big]\big[\id s+\mB^{(k|i)}+\mB^{(i|k)}-\mL^{k,i}\big]^{-1}\bc^{(k|i)}
}
 .
 \label{eq:twp}
\end{align}

The probability that starting from the observed state $i$ the next state is going to be $j$
is given by the splitting probability, $ \phi_{j|i}=\int_0^\infty\wp_{j|i}(t)\dd t$, which explicitly reads
\begin{equation}
 \phi_{j|i}=\twp_{j|i}(0)
 =\frac{ w_{i\to j}+(\bbb^{(i|j)})^\T\big[\mB^{(j|i)}+\mB^{(i|j)}-\mL^{i,j}\big]^{-1}\bc^{(j|i)}}{r^{(i)} +\sum_k(\bbb^{(i|k)})^\T\big[\mB^{(k|i)}+\mB^{(i|k)}-\mL^{k,i}\big]^{-1}\bc^{(k|i)}}.
 \label{eq:splitting}
\end{equation}
As soon as there are no hidden states (e.g., $\bbb^{(j|i)}=0$) rendering the observed dynamics becomes Markovian with $ \phi_{j|i}= w_{i\to j}/ r^{(i)}$ and $\twp_{j|i}(s)/\phi_{j|i}=r^{(i)}/(s+r^{(i)})$, which is the Laplace transform of an exponential density $\wp_{j|i}(t)/\phi_{j|i}=r^{(i)}\exp(-r^{(i)} t)$. 

In the following, we characterize the hidden
dynamics and the emergence of observed
non-Markovian dynamics.

\begin{figure}[ht]
\centering
\includegraphics[width=0.75\textwidth]{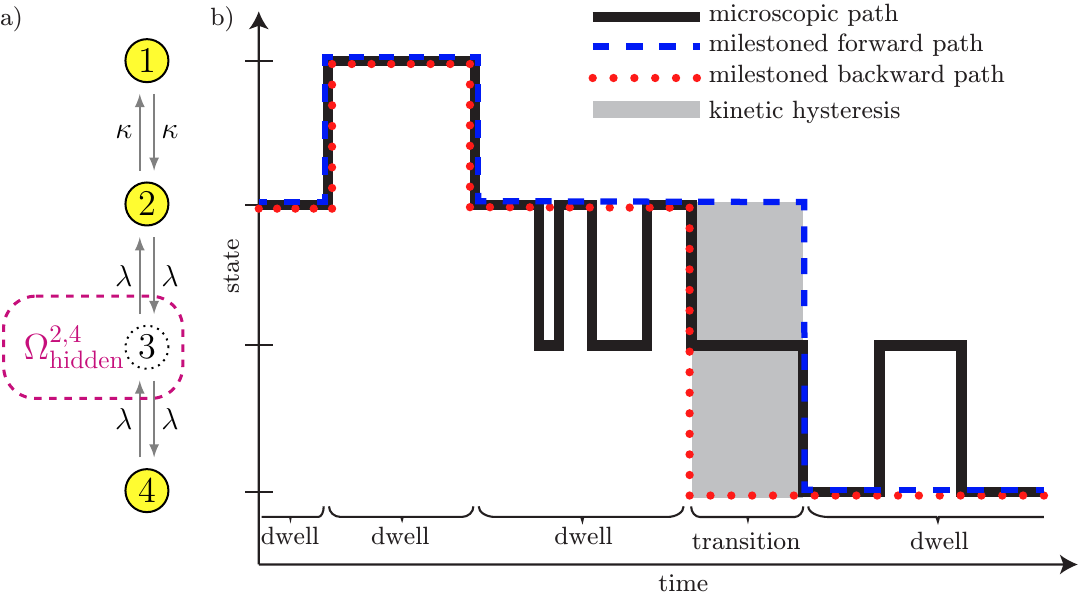}
 \caption{A milestoned Markov process and its transition path with kinetic hysteresis: a) Microscopic model with three milestones (1,2,4) and one hidden state (3). b) Microscopic path (black) and its forward milestones (dashed blue) and backward milestoned (dotted red) path. While the waiting time of the mesoscopic $1 \rightarrow 2$ transition consists only of dwell time (left) and is therefore identical under time reversal, the waiting time of the mesoscopic $2\rightarrow4$ transition contains non-zero transition time (right), leading to kinetic hysteresis (gray area is non-zero). Importantly, even under time-scale separation $\lambda \gg \kappa$ which leads to approximately Markovian dynamics (waiting times are approximately exponential), the effect of kinetic hysteresis on the stochastic thermodynamics persists (see Section~\ref{sec:kin-hyst}).
 }
 
 \label{fig:time-scale-sep-but-kinetic-hysteresis}
\end{figure}

\subsubsection{Invariant measure and current for coarse-grained dynamics}Since we work with an ergodic system, we will observe in the long time limit ($t\rightarrow \infty$) many repeated jumps with average effective transition rates\footnote{For a detailed discussion, see Appendix~\ref{app:effecitve-rates-ergodic}.}

\begin{align}
w_{i \rightarrow j}^{\rm eff} = \frac{\phi_{j \mid i}}{\int_0^\infty t \wp^{\rm exit}_i(t) \dd t },\quad \text{for}\quad i,j \in \Omega_{\rm obs}, \label{eq:effective-rates}
\end{align}
where the mean exit time follows from the Laplace transform of Eq.~\eqref{eq:exit_pre} as
\begin{align}
     \int_0^\infty t\wp^{\rm exit}_i(t) \dd t = 
     \left(
 r^{(i)},(\bbb^{(i|1)})^\T,\dots,(\bbb^{(i|N)})^\T\right)
\Big[\mL^{(i)}\Big]^{-2}
 \boldsymbol{e}_1. 
\end{align}
The invariant distribution $\boldsymbol{p}^{\rm inv}$ on $\Omega_{\rm obs}$ corresponds to the normalised right eigenvector of eigenvalue $0$ of the corresponding transition rate matrix with off-diagonal elements $(\mathbf{L}^{\rm eff})_{ji}=w_{i\rightarrow j}^{\rm eff}$ and diagonals $(\mathbf{L}^{\rm eff})_{ii}=-\sum_{j \neq i} w_{i\rightarrow j}^{\rm eff}$.

The average number of transitions from milestone $i$ to milestone $j$ per unit time in the steady state is furthermore given by
\begin{align}
   \dot n_{j \mid i} = \frac{p^{\rm inv}_i \phi_{j \mid i} }{\left(
 r^{(i)},(\bbb^{(i|1)})^\T,\dots,(\bbb^{(i|N)})^\T\right)
\Big[\mL^{(i)}\Big]^{-2}
 \boldsymbol{e}_1},
 \label{ndot}
\end{align}
which fully characterizes the stationary kinetic properties of the milestoned dynamics. 

\subsubsection{Dissecting waiting time into dwell and transition time (flavor of memory)}
As soon as there are hidden states connecting two observed nodes the dynamics
develops a memory and the statistics of the waiting time Eq.~\eqref{eq:twp} deviates from an exponential distribution, which in Laplace space means $\twp_{j|i}(s)\neq\phi_{j|i}r^{(i)}/(s+r^{(i)})$.
In the following we show that waiting time density can be decomposed into
the statistics of transition time and the remaining dwell time statistics (see Fig.~\ref{fig:time-scale-sep-but-kinetic-hysteresis} for an example). This decomposition is the key towards an understanding of the emergence of non-Markovian dynamics.

The \emph{transition time} 
within a state change  $i\to j$  denotes the period between the last departure
from state $i$ at time and the first entrance into state $j$. Any finite transition time immediately triggers a memory in the observed dynamics. Inspecting the transition probability Eq.~\eqref{eq:splitting} we find the probability can be decomposed into $\phi_{j|i}=\phi_{j|i}^{\rm jump}+\phi_{j|i}^{\rm hidden}$ with
\begin{equation}
\begin{aligned}
  \phi_{j|i}^{\rm jump}
 &=\frac{ w_{i\to j}}{r^{(i)} +\sum_k(\bbb^{(i|k)})^\T\big[\mB^{(k|i)}+\mB^{(i|k)}-\mL^{k,i}\big]^{-1}\bc^{(k|i)}},
 \\
   \phi_{j|i}^{\rm hidden}
 &=\frac{ (\bbb^{(i|j)})^\T\big[\mB^{(j|i)}+\mB^{(i|j)}-\mL^{i,j}\big]^{-1}\bc^{(j|i)}
 }{
 r^{(i)} +\sum_k(\bbb^{(i|k)})^\T\big[\mB^{(k|i)}+\mB^{(i|k)}-\mL^{k,i}\big]^{-1}\bc^{(k|i)}
 }.
 \label{eq:splitting_dec}
\end{aligned}
\end{equation}
The normalized statistics of transition time is given by
\begin{equation}
 \wp^{\rm {tr, full}}_{j|i}(t)=\phi_{j|i}^{\rm jump}\delta (t)+\phi_{j|i}^{\rm hidden}\psi^{\rm tr}_{j|i}(t)
 \label{eq:wptrans}
\end{equation}
with $\int_0^\infty \delta (t)\dd t=\int_0^\infty \psi^{\rm tr}_{j|i}(t)\dd t=1$, where $\delta(t)$ denotes the Delta-distribution for instantaneous jumps, and where $\psi^{\rm tr}_{j|i}$ needs to be determined in the following.

The transition $i\to j$ starts by leaving the state $i$, and
each state within $\Omega^{(j|i)}_{\rm hidden}$ is initially 
entered with probability vector $\bc^{(j|i)}/\|\bc^{(j|i)}\|_1 $.
The probability density to 
complete the transition at time $t$
is
\begin{equation}
 \psi^{\rm tr}_{j|i}(t)=K^{-1}\bone \mB^{(i|j)}\exp\left[(\mL^{i,j}-\mB^{(j|i)}-\mB^{(i|j)})t\right]\frac{\bc^{(j|i)}}{\|\bc^{(j|i)}\|_1},
 \label{eq:ptrans_pre}
\end{equation}
where $K$  guarantees normalization $ \int_0^\infty\psi^{\rm tr}_{j|i}(t)\dd t=1$.
Note that  $K$ is  the probability of a successful completion of a transition after entering the hidden network $\Omega^{(j|i)}_{\rm hidden}$ from $i$ to $j$, whereas $1-K$ is the probability of a recursion from $i$ back to $i$, i.e., a failed transition attempt. Using $\bone \,\mB^{(i|j)}=(\bbb^{(i|j)})^\T$ and Eq.~\eqref{eq:ptrans_pre} the probability density of the indirect transition becomes
\begin{equation}
  \psi^{\rm tr}_{j|i}(t)=\frac{(\bbb^{(i|j)})^\T\big[\exp\left((\mL^{i,j}-\mB^{(j|i)}-\mB^{(i|j)})t\right)\big]\bc^{(j|i)}}{(\bbb^{(i|j)})^\T\big[\mB^{(j|i)}+\mB^{(i|j)}-\mL^{i,j}\big]^{-1}\bc^{(j|i)}}
  \label{eq:ptrans}
\end{equation}
along with its Laplace transform
\begin{equation}
  \tilde\psi^{\rm tr}_{j|i}(s)=\frac{(\bbb^{(i|j)})^\T\big[\id s+\mB^{(j|i)}+\mB^{(i|j)}-\mL^{i,j}\big]^{-1}\bc^{(j|i)}}{(\bbb^{(i|j)})^\T\big[\mB^{(j|i)}+\mB^{(i|j)}-\mL^{i,j}\big]^{-1}\bc^{(j|i)}}.
  \label{eq:tptrans}
\end{equation}

The Laplace transform of the full transition time from Eq.~\eqref{eq:wptrans} which includes both direct instantaneous \emph{and} indirect  non-instantaneous transitions reads
\begin{equation}
\twp^{\rm {tr, full}}_{j|i}(s)
=
\frac{w_{i\to j}+ (\bbb^{(i|j)})^\T\big[\id s+\mB^{(j|i)}+\mB^{(i|j)}-\mL^{i,j}\big]^{-1}\bc^{(j|i)}}{
r^{(i)} +\sum_k(\bbb^{(i|k)})^\T\big[\mB^{(k|i)}+\mB^{(i|k)}-\mL^{k,i}\big]^{-1}\bc^{(k|i)}
}
,
\label{eq:twptrans}
\end{equation}
where we used Eqs.~\eqref{eq:splitting_dec} and \eqref{eq:tptrans} as well as $\int_0^\infty\exp(-st)\delta (t)\dd t=1$.

The \emph{dwell time} period is independent of the transition time and therefore can be deduced from
\begin{align}
 \twp^{\rm dwell}_{j|i}(s)&=\frac{\twp_{j|i}(s)}{\twp^{\rm {tr, full}}_{j|i}(s)}\nonumber\\
 &=
\frac{
r^{(i)} +\sum_k(\bbb^{(i|k)})^\T\big[\mB^{(k|i)}+\mB^{(i|k)}-\mL^{k,i}\big]^{-1}\bc^{(k|i)}
}{
s+r^{(i)} +\sum_k\big[s \bone+(\bbb^{(i|k)})^\T\big]\big[\id s+\mB^{(k|i)}+\mB^{(i|k)}-\mL^{k,i}\big]^{-1}\bc^{(k|i)}
},
\label{eq:twp_dwell}
\end{align}

Notably, the dwell time does not depend on the final state $j$. The decomposition of the waiting time distribution into transition and dwell time is the main result of this paper. This extends the result which we deduced for diffusive dynamics to discrete state jump processes, which includes both direct and indirect transitions. In the following we address both  kinetics and thermodynamics on the dynamics of the coarse-graining.

\subsection{Symmetries of coarse-grained dynamics}

\subsubsection{Thermodynamic consistency}
From Eqs.~\eqref{eq:quasiLDB} and \eqref{eq:splitting_dec}, it follows immediately $ \phi_{j|i}^{\rm jump}/ \phi_{i|j}^{\rm jump}=\exp(\mathcal{A}_{ij})$. We will now show that  $ \phi_{j|i}^{\rm hidden}/ \phi_{i|j}^{\rm hidden}=\exp(\mathcal{A}_{ij})$, which in turn renders the coarse-graining to be thermodynamically consistent in the sense that it preserves the steady-state entropy production \cite{schn76} (see also \cite{wang07,hartichEmergentMemoryKinetic2021a}). From Eq.~\eqref{eq:quasiLDB} follows that $\mL^{i,j}$ satisfies detailed balance \cite{kamp07}, that is $L^{j,i}_{\beta \alpha}\pi^{(j|i)}_\alpha=L^{j,i}_{ \alpha\beta}\pi^{(j|i)}_\beta$, where $\boldsymbol{\pi}^{(j|i)}$ is the ``equilibrium distribution of $\mL^{i,j}$ alone''. This implies the following relations \cite{kamp07}
\begin{equation}
(\mL^{i,j})^{\T}= \operatorname{diag}[\boldsymbol{\pi}^{(j|i)}]^{-1}\mL^{i,j} \operatorname{diag}[\boldsymbol{\pi}^{(j|i)}]
\label{eq:DB1}
\end{equation}
and
\begin{equation}
\exp(\mathcal{A}_{ij})=
\frac{w_{i\to \alpha}}{ w_{\alpha\to i}} \times\frac{\pi^{(j|i)}_\beta}{\pi^{(j|i)}_\alpha}\times \frac{w_{\beta\to j}}{ w_{j \to \beta}}=\frac{c^{(j|i)}_\alpha}{ b^{(j|i)}_\alpha} \times\frac{\pi^{(j|i)}_\beta}{\pi^{(j|i)}_\alpha}\times \frac{b^{(i|j)}_\beta}{ c^{(i|j)}_\beta},
\label{eq:DB2}
\end{equation}
where in the last step we identified  Eqs.~\eqref{eq:hidden>obs_vec} and \eqref{eq:obs>hidden_vec}. Eq.~\eqref{eq:DB2} can be rewritten in the form of the matrix product
\begin{equation}
 (\bbb^{(i|j)})^\T[\cdots]  \bc^{(j|i)}= \exp(\mathcal{A}_{ij})(\bc^{(i|j)})^\T\operatorname{diag}[\boldsymbol{\pi}^{(j|i)}]^{-1}[\cdots] \operatorname{diag}[\boldsymbol{\pi}^{(j|i)}]\bbb^{(j|i)},
 \label{eq:DB3}
\end{equation}
which in conjunction with the second line of Eq.~\eqref{eq:splitting_dec}
yields
\begin{equation}
\phi_{j|i}^{\rm hidden}=\phi_{i|j}^{\rm hidden}\frac{R_j}{R_i}\exp(\mathcal{A}_{ij}),
\end{equation}
where we in addition used  $(\bbb^{(i|j)})^\T[\cdots]  \bc^{(j|i)}=(\bc^{(j|i)})^\T[\cdots]^\T  \bbb^{(i|j)}$ and finally used Eq.~\eqref{eq:DB1} as well as $R_i\equiv r^{(i)} +\sum_k(\bbb^{(i|k)})^\T\big[\mB^{(k|i)}+\mB^{(i|k)}-\mL^{k,i}\big]^{-1}\bc^{(k|i)}$. This completes the proof that \eqref{eq:quasiLDB} implies
\begin{align}
\ln (\phi_{j|i}/\phi_{i|j})= \ln (\phi_{j|i}^{\rm jump}/\phi_{i|j}^{\rm jump})= \ln (\phi_{j|i}^{\rm hidden}/\phi_{i|j}^{\rm hidden}) =\mathcal{A}_{ij}+\ln R_j-\ln R_i.
\label{eq:LDBgen}
\end{align}
Denoting the number of transitions from observed node $i$ to $j$ per unit time as \cite{wang07}
\begin{equation}
 \dot n_{j|i}=\lim_{t\to\infty} \frac{\text{number of observed transitions $i$ to $j$ until time $t$}}{t},
\end{equation}
which we determined explicitly in Eq.~\eqref{ndot} above,
the total dissipation becomes \cite{schn76}
\begin{equation}
\dot \sigma\equiv\frac{1}{2}\sum_{\alpha,\beta\in\Omega} \big[\pss_\alpha w_{ \alpha\to \beta}-\pss_\beta w_{ \beta\to \alpha}\big]\ln\frac{w_{\alpha\to\beta}}{w_{\beta\to\alpha}}=\frac{1}{2}\sum_{i,j\in \Omega_{\rm obs}}\big[\dot n_{j|i}-\dot n_{i|j}\big]\mathcal{A}_{ij}.
\label{eq:entropy_def}
\end{equation}
If we now use Eq.~\eqref{eq:LDBgen} and Kirchhoff's law $\sum_j \dot n_{j|i}=\sum_j\dot n_{i|j}$ the dissipation \eqref{eq:entropy_def}
becomes
\begin{equation}
 \dot \sigma=\frac{1}{2}\sum_{i,j\in \Omega_{\rm obs}}\big[\dot n_{j|i}-\dot n_{i|j}\big]\big(\mathcal{A}_{ij}+\ln R_j-\ln R_i\big)=\sum_{i,j\in \Omega_{\rm obs}}\dot n_{j|i}\ln\frac{\phi_{j|i}}{\phi_{i|j}}.
 \label{eq:entropy_preservation}
\end{equation}
The coarse-graining preserves the steady-state entropy production \eqref{eq:entropy_preservation}
since in Eq.~\eqref{eq:quasiLDB} we do not hide away dissipative cycles (see also Ref.~\cite{pugl10}). We note that this readily confirms that splitting probabilities (affinities) \emph{fully} encode the dissipation of semi-Markov dynamics also for finite duration transitions \cite{hartichEmergentMemoryKinetic2021a}. In contrast to our previous  work \cite{hartichEmergentMemoryKinetic2021a} we here consider the observed non-Markovian renewal dynamics to be generated by an underlying Markov-jump process.  

If one assumes a naive definition of the steady-state entropy production rate for the renewal process characterized by states $i$ and waiting times $\tau_i$ based on the violation of ``mathematical'' time-reversal symmetry \cite{mart19a}, i.e.\ a relative entropy between forward paths $\{i,\tau_i\}_{0\le i\le N}$ and backward paths $\{N-i,\tau_{N-i}\}_{0\le i\le N}$ with $\sum_{i=0}^N\tau_i=T$, the result yields two contributions \cite{Girardin_2003,mart19a}
\begin{align}
\!\!\!\!\! D_{\rm KL}[\mathbb{P}(\{i,\tau_i\}_{0\le i\le N}||\mathbb{P}(\{N-i,\tau_{N-i}\}_{0\le i\le N})]=\dot{\sigma} + \sum_{i,j,k} \phi_{k|j}\dot{n}_{j|i}D_{\rm KL}[\phi^{-1}_{k|j}\wp_{k|j}||\phi^{-1}_{i|j}\wp_{i|j}],
\label{KL_all}    
\end{align}
whereby the second, so-called waiting-time contribution is also strictly non-negative. 
Note, therefore, that as soon as the transition paths become both, finite (i.e., non-instantaneous) and unequally fast (e.g., $\phi^{-1}_{k|j}\wp_{k|j}\ne \phi^{-1}_{l|j}\wp_{l|j}$ for some $k\ne l$)
Eqs.~\eqref{eq:entropy_def} and \eqref{eq:entropy_preservation} contradict the finding of Ref.~\cite{mart19a}. To the best of our knowledge this contradiction was first found in Ref.~\cite{wang07}, but not explicitly attributed to transition paths (for a further discussion see \cite{hartichCommentInferringBroken2024}).

\subsubsection{Reflection symmetry of transition-path times} \label{sec:symm-transition-path-times}

It is nicely shown in Ref.~\cite{bere19a} that the transition time statistics obeys a reflection symmetry $ \psi^{\rm tr}_{j|i}(t)= \psi^{\rm tr}_{i|j}(t)$. This symmetry holds as soon as the dynamics within the transition region, Eq.~\eqref{eq:L_hidden}, locally satisfies detailed balance. In the following paragraph we briefly reconfirm the proof in Ref.~\cite{bere19a}. Then we show that the proof extends also to the case of possible instantaneous transitions (see also Ref.~\cite{bere06} for a diffusion approach).

Let us first focus on the proof that the indirect transitions obey a forward/backward symmetry $ \tilde \psi^{\rm tr}_{j|i}(s)= \tilde\psi^{\rm tr}_{i|j}(s)$. The corresponding (indirect) backward transition  Eq.~\eqref{eq:ptrans} reads
\begin{equation}
  \tilde\psi^{\rm tr}_{i|j}(s)=\frac{(\bbb^{(j|i)})^\T\big[\id s+\mB^{(j|i)}+\mB^{(i|j)}-\mL^{i,j}\big]^{-1}\bc^{(i|j)}}{(\bbb^{(j|i)})^\T\big[\mB^{(j|i)}+\mB^{(i|j)}-\mL^{i,j}\big]^{-1}\bc^{(i|j)}},
  \label{eq:tptrans_backward1}
\end{equation}
where $\mL^{i,j}=\mL^{i,j}$ holds by definition \eqref{eq:L_hidden}. Inserting Eqs.~\eqref{eq:DB1} and \eqref{eq:DB3} into \eqref{eq:tptrans_backward1} we obtain
\begin{align}
  \tilde\psi^{\rm tr}_{i|j}(s)
  =
  \frac{
  (\bc^{(j|i)})^\T(\big[\id s+\mB^{(j|i)}+\mB^{(i|j)}-(\mL^{i,j})^\T\big]^{-1}\bbb^{(i|j)}
  }{
(\bc^{(j|i)})^\T\big[\mB^{(j|i)}+\mB^{(i|j)}-(\mL^{i,j})^\T\big]^{-1}\bbb^{(i|j)}}=\tilde\psi^{\rm tr}_{j|i}(s),
  \label{eq:tptrans_backward2}
\end{align}
where in the last step we used the fact that $\mB^{(j|i)}=(\mB^{(j|i)})^\T$ are diagonal matrices as well as that transposing  the numerator and denominator does not affect the fraction since both are  $1\times1$ matrices (scalars). This completes the proof of $\psi^{\rm tr}_{i|j}=\psi^{\rm tr}_{j|i}$ (see Ref.~\cite{bere19a} for an alternative proof). 

Let us now incorporate instantaneous transitions. The full density in Eq.~\eqref{eq:wptrans}
satisfies $\int_0^\infty \wp^{\rm {tr, full}}_{j|i}(t)\dd t=\phi_{j|i}$.
The normalized backward transition $j\to i$
is distributed with the density
\begin{equation}
 \frac{\wp^{\rm {tr, full}}_{j|i}(t)}{\phi_{j|i}}=\frac{\phi_{j|i}^{\rm jump}\delta (t)+\phi_{j|i}^{\rm hidden}\psi^{\rm tr}_{j|i}(t)}{\phi_{j|i}}.
 \label{eq:wptrans_full_back1}
\end{equation}
Inserting Eq.~\eqref{eq:LDBgen} and $\psi^{\rm tr}_{j|i}(t)=\psi^{\rm tr}_{i|j}(t)$
into Eq.~\eqref{eq:wptrans_full_back1}
finally yields
\begin{equation}
 \frac{\wp^{\rm {tr, full}}_{j|i}(t)}{\phi_{j|i}}=\frac{\phi_{j|i}^{\rm jump}\delta (t)+\phi_{j|i}^{\rm hidden}\psi^{\rm tr}_{j|i}(t)}{\phi_{j|i}}=\frac{\wp^{\rm {tr, full}}_{i|j}(t)}{\phi_{i|j}}.
 \label{eq:wptrans_full_back2}
\end{equation}
Note that in the first step in Eq.~\eqref{eq:wptrans_full_back2} we explicitly used the identity  $\phi_{i|j}^{\rm jump}/\phi_{i|j}=(\phi_{j|i}^{\rm jump}/\phi_{j|i})\times [(\phi_{i|j}^{\rm jump}/\phi_{j|i}^{\rm jump})/(\phi_{i|j}/\phi_{j|i})]$ while  Eq.~\eqref{eq:LDBgen} implies that the bracketed term satisfies $[\cdots]=1$. Eq.~\eqref{eq:wptrans_full_back2} ``extends'' the forward/backward symmetry shown in Ref.~\cite{bere19a} by explicitly including possibly \emph{direct} instantaneous transitions.

\subsubsection{Summarizing all three symmetries in the time domain}
Let us denote the waiting time by $t$, which is the independent sum of the dwell time $\tau$ and transition time $\delta t=t-\tau$.
The joint waiting time density $\wp_{j|i}(t)$ thus is a convolution
\begin{align}
 \wp_{j|i}(t)&=\int_0^t\wp^{\rm {tr, full}}_{j|i} (t-\tau) \wp^{\rm dwell}_{i}(\tau)\dd \tau
 =\phi_{j|i}^{\rm jump}\wp^{\rm dwell}_{i}(t)+\phi_{j|i}^{\rm hidden}\int_0^t\psi^{\rm {tr}}_{j|i} (t-\tau) \wp^{\rm dwell}_{i}(\tau)\dd \tau,
 \label{eq:independence}
\end{align}
where we defined $\wp^{\rm dwell}_{i}\equiv \wp^{\rm dwell}_{j|i}$, since the dwell time statistics in  Eq.~\eqref{eq:twp_dwell} does not depend on the final state $j$, which we call \emph{symmetry 1}. Symmetry 1 means that the dwell time statistics is a property of an \emph{observed state} $i$. The reflection identity of the transition time statistics Eq.~\eqref{eq:wptrans_full_back2} means that the transition time statistics is a property of the \emph{link connecting a pair of observed states $i$ and $j$} which we in the following call \emph{symmetry 2}. Symmetry 2 entails the reflection identity $\psi^{\rm {tr}}_{j|i}=\psi^{\rm {tr}}_{i|j}$ proven in \cite{bere19a} and reaffirmed in Eq.~\eqref{eq:tptrans_backward2}.
It is worth mentioning that the independence between dwell time and transition time, Eq.~\eqref{eq:independence}, separates these two individual ``flavors'' of the memory \cite{hartichEmergentMemoryKinetic2021a}.
\emph{Symmetry 3} is listed in Eq.~\eqref{eq:LDBgen} and relates the splitting probability to the dissipation and guarantees that the entropy production is preserved [cf. Eqs.~\eqref{eq:entropy_def} and \eqref{eq:entropy_preservation}]. 

\subsubsection{Manifestation of kinetic hysteresis, even under time-scale separation} \label{sec:kin-hyst}
Whenever $\wp^{\rm {tr, full}}_{j|i}(t) \neq \delta(t)$,
the coarse-graining does not commute with 
the time-reversal which leads to what we call \emph{kinetic hysteresis} \cite{hartichEmergentMemoryKinetic2021a}. The kinetic hysteresis prevents the
waiting time distribution from encoding irreversible non-Markovian dynamics while satisfying
detailed balance as found in Ref.~\cite{wang07}. Importantly, kinetic hysteresis may even occur when the microscopic dynamics obeys detailed balance \cite{hartichCommentInferringBroken2024,hartichEmergentMemoryKinetic2021a}.
Even further, the presence of time-scale separation leading to approximately Markov (to a high degree) kinetics does not remove the impact of the kinetic hysteresis on the stochastic thermodynamics, as the following example shows.

\begin{figure}
 \includegraphics[width=0.8\textwidth]{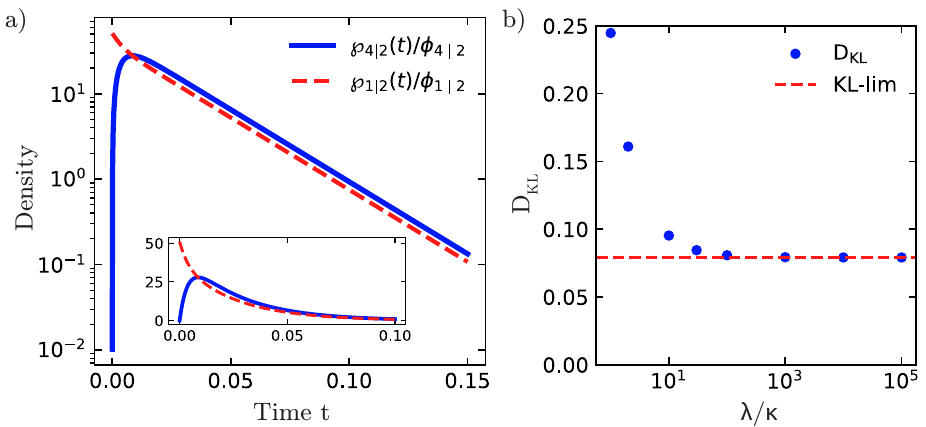}
  \caption{Manifestation of kinetic hysteresis even in the presence of time-scale separation, i.e., in the Markovian limit. (a)~Waiting time densities become almost exponential in the presence of a time-scale separation (i.e., fast hidden dynamics) in Fig.~\ref{fig:time-scale-sep-but-kinetic-hysteresis}: For $\kappa\equiv 1$, $\lambda \equiv 100$, we converge to Markov dynamics. (b)~Convergence of $D_{\mathrm{KL}}\left({\wp_{4|2}(t)}/{\phi_{4 \mid 2}} || {\wp_{1|2}(t)}/{\phi_{1 \mid 2}} \right)$ to a non-zero value even under time-scale separation.}
 \label{fig:kinetic-hysteresis-but-markov}
\end{figure}

We consider the system depicted in Fig.~\ref{fig:time-scale-sep-but-kinetic-hysteresis} with $\lambda \gg \kappa$. The waiting time density $2\to4$ is given by Eq.~\eqref{eq:wp}, i.e.\
\begin{align}
 \wp_{4|2}(t)&=
\left(0,(\bbb^{(2|4)})^\T\right)
\exp\left(\mL^{(2)}t\right)
 \boldsymbol{e}_1 \\
 &= \left(0,(\bbb^{(2|4)})^\T\right)
\exp\left(\begin{pmatrix}
    -\kappa -\lambda & \lambda \\
    \lambda & -2 \lambda
\end{pmatrix}t\right)
 \boldsymbol{e}_1,
\end{align}
while for the transition $2\to 1$ we find
\begin{align}
 \wp_{1|2}(t)&=
\left(w_{2 \rightarrow 1},0\right)
\exp\left(\mL^{(2)}t\right)
 \boldsymbol{e}_1.
\end{align}
However, even in the limit of divergingly fast hidden dynamics the KL divergence between the (normalised) waiting time densities remains non-zero
\begin{align}
    D_{\mathrm{KL}}\left(\phi^{-1}_{4 \mid 2}\wp_{4|2}(t)||\phi^{-1}_{1 \mid 2}\wp_{1|2}(t)\right) > 0.
\end{align}
More precisely, we have
\begin{align}
    \frac{\wp_{4|2}(t)}{\phi_{4 \mid 2}}&= \frac{2\lambda (1+\frac{2 \kappa }{\lambda})}{\sqrt{5-\frac{2\kappa}{\lambda}+\frac{\kappa^2}{\lambda^2}}} \exp\left[-\frac{1}{2}\lambda t\left(3+\frac{\kappa}{\lambda} \right) \right] \sinh\left[\frac{1}{2}\lambda t \sqrt{5-\frac{2 \kappa}{\lambda}+\frac{\kappa^2}{\lambda^2}} \right].
\end{align}
and 
\begin{align}
    \frac{\wp_{1|2}(t)}{\phi_{1 \mid 2}}&=\frac{1}{2}\lambda \left(1+ \frac{2\kappa}{\lambda} \right)\exp \left[-\frac{1}{2}\lambda t\left(3+\frac{\kappa}{\lambda} \right) \right] \Bigg\{ \cosh \left[ \frac{1}{2}\lambda t \sqrt{5- \frac{2\kappa}{\lambda} + \frac{\kappa^2}{\lambda^2}} \right]  \\
    & \qquad
    +\frac{1- \frac{\kappa}{\lambda}}{\sqrt{5-\frac{2\kappa}{\lambda} + \frac{\kappa^2}{\lambda^2}}}  \sinh \left[ \frac{1}{2}\lambda t \sqrt{5- \frac{2\kappa}{\lambda} + \frac{\kappa^2}{\lambda^2}} \right] \Bigg\} .
\end{align}
so that the KL divergence converges in the scale limit $\lambda/\kappa  \rightarrow \infty$ to
\begin{align}
 \!\! & \lim_{\lambda/\kappa \rightarrow \infty }D_{\mathrm{KL}}\left(\phi^{-1}_{4 \mid 2}\wp_{4|2}(t)||\phi^{-1}_{1 \mid 2}\wp_{1|2}(t)\right)  \\ 
  =& \int_0^\infty \frac{2}{\sqrt{5}} \exp\left( -\frac{3}{2} y\right) \ln \left[ \frac{4}{1+\sqrt{5}\coth\left(\frac{\sqrt{5}}{2} y \right)} \right]\sinh\left( \frac{\sqrt{5}}{2}y\right) \dd y
  \approx 0.079 > 0.
\end{align}
In Fig.~\ref{fig:kinetic-hysteresis-but-markov} we show the normalized conditioned waiting-time-densities for transitions $2\to1$ and $2\to 4$ for $\kappa=1$ and $\lambda = 100$ (panel a) as well as $D_{\mathrm{KL}}\left(\phi^{-1}_{4 \mid 2}\wp_{4|2}(t)||\phi^{-1}_{1 \mid 2}\wp_{1|2}(t)\right)$ (panel b)  for different values of $\lambda \gg \kappa=1$. In particular, it can be seen in Fig.~\ref{fig:kinetic-hysteresis-but-markov}a that the waiting time densities for $\lambda = 100$ become approximately exponential whereas $D_{\mathrm{KL}}$ converges to a strictly positive constant. Since the system is manifestly in thermodynamic equilibrium, the entropy production rate must be $\dot S \equiv 0$.

Based on the symmetry in Eq.~\eqref{eq:independence} we find that kinetic hysteresis emerges if and only if there exist $i,j,k$ such that $\wp^{\rm {tr, full}}_{j|i}(t)\ne \wp^{\rm {tr, full}}_{k|i}(t)$. This in turn implies that $D_{\mathrm{KL}}\left({\wp_{j|i}(t)}/{\phi_{j \mid i}} || {\wp_{k|i}(t)}/{\phi_{k \mid i}} \right) > 0$ if and only if we have kinetic hysteresis. The presence of kinetic hysteresis in semi-Markov dynamics is completely unrelated to the (non)existence of detailed balance in microscopic Markov dynamics. 

Therefore, in the presence of kinetic hysteresis a definition of dissipation 
based on the naive ``mathematical'' time-reversal of the coarse-graining does \emph{not} encode dissipation
\begin{align}
\lim_{T\to\infty}\frac{1}{T} D_{\rm KL}[\mathbb{P}(\{i,\tau_i\}_{0\le i\le N}||\mathbb{P}(\{N-i,\tau_{N-i}\}_{0\le i\le N})] > \dS,
\label{violation}
\end{align}
where the constant $T=\sum_{i}\tau_i$ is the duration of the observation.
Remarkably, the above example shows that even in the limit of Markovian coarse-grained kinetics in the presence of  time-scale separation in the hidden dynamics, \emph{kinetic hysteresis} causes the unphysical result Eq.~\eqref{violation}. 

The ``complication'' caused by kinetic hysteresis may be resolved in two \emph{equivalent} ways: either as transition-path times being ``odd'' under time-reversal, which prescribes the correct form of physical time-reversal operation with accounting for waiting times for semi-Markov processes emerging from Markov dynamics (and forces the ``waiting-time'' contribution $\dS$ to zero), or, alternatively, that the waiting times are simply ignored in defining physical time-reversal (see also \cite{hartichEmergentMemoryKinetic2021a,hartichCommentInferringBroken2024}).

\newcommand{\Niminusj}{{N(i) \setminus \{j\}}}
\newcommand{\wptext}[1]{\wp_\text{#1}}
\newcommand{\tildewptext}[1]{\tilde \wp_\text{#1}}

\subsection{Generalization of the renewal network theory}
\label{sec:generalization}
\subsubsection{Parallel mechanisms}
In Eq.~\eqref{eq:quasiLDB} we only assume \emph{one} mechanism that yields a mesoscopic state change $i\to j$. It is tedious but conceptually straightforward to incorporate multiple parallel mechanisms $\nu=1,\ldots,\mathcal{V}$ as in Ref.~\cite{espo08}. First, we partition the hidden states into $\Omega_{\rm hidden}^{i,j}=\bigcupdot_{\nu=1}^\mathcal{V}\Omega_{\rm hidden}^{i,j}(\nu)$ (see also Ref.~\cite{Esposito2012PRE} for an instructive presentation). The $\Omega_{\rm hidden}^{i,j}(\nu)$ connects observed states $i$ and $j$ via mechanism $\nu$
and is assumed to be disconnected from other mechanisms, i.e.\  $\Omega_{\rm hidden}^{i,j}(\nu)\cap \Omega_{\rm hidden}^{i,j}(\nu')=\emptyset$ for all $\nu\neq\nu'$. Direct transitions via mechanism $\nu$ occur with rate $w^{(\nu)}_{i\to j}$ such that the total rate of direct jumps $i\to j$ becomes $w_{i\to j}=\sum_\nu w^{(\nu)}_{i\to j}$. Each block in the matrix Eq.~\eqref{eq:generator} separates into distinct blocks containing the various mechanisms.

In this case all paths $(x_0\to x_1\to\ldots\to x_n)$ starting from $x_0=i$ to state $x_n=j$ $(n\ge1)$ along one mechanism $\nu$, i.e., paths on
$x_ \tau\in\Omega_{\rm hidden}^{i,j}(\nu)\cup\{i,j\}$, satisfy
\begin{equation}
\sum_{\tau=0}^{n-1}\ln\frac{w^{(\nu)}_{x_{\tau}\to x_{\tau+1}}}{w^{(\nu)}_{x_{\tau+1}\to x_{\tau}}}=\mathcal{A}_{ij}(\nu).
\label{eq:quasiLDB_mult}
\end{equation}
One can show that all quantities  entering Eq.~\eqref{eq:LDBgen} (except $R_i,R_j$) become functions of the mechanism $\nu$.
The entropy production \eqref{eq:entropy_preservation} becomes
\begin{equation}
\dot \sigma=\sum_{\nu=1}^\mathcal{V}
\sum_{i,j\in \Omega_{\rm obs}}\dot n_{j|i}(\nu)\ln\frac{\phi_{j|i}(\nu)}{\phi_{i|j}(\nu)}
\ge \sum_{i,j\in \Omega_{\rm obs}}\dot n_{j|i}\ln\frac{\phi_{j|i}}{\phi_{i|j}},
\label{eq:entropy_multichannel}
\end{equation}
where $\dot n_{j|i}(\nu)$ is the transition rate $i\to  j$ along mechanism $\nu$ and $\phi_{j|i}(\nu)$ the corresponding splitting probability. The last inequality in Eq.~\eqref{eq:entropy_multichannel} assumes $\phi_{j|i}\equiv\sum_\nu \phi_{j|i}(\nu)$ and  $\dot{n}_{j|i}\equiv\sum_\nu \dot{n}_{j|i}(\nu)$.
In general the reflection symmetry Eq.~\eqref{eq:wptrans_full_back2} ceases to hold, whereas along
the  individual mechanisms the reflection symmetry remains. More precisely,
the transition time statistics $i\to j$ along mechanism $\nu$ is \emph{always} the same as the backward 
transition time statistics $j\to i$ along the \emph{same} mechanism $\nu$.

Along distinct mechanisms the symmetry is expected to break down
as shown experimentally in Ref.~\cite{glad19}, where
it was found that a tagged particle in single file
transits faster uphill than downhill \cite{ryab19}. This quite counterintuitive finding can be attributed to distinct physical transport mechanisms \cite{ryab19}. For example, while downhill the motion is steadily pushed by other particles, the uphill transition is hindered by other moving particles and thus leads to the quite counterintuitive phenomenon.

\subsubsection{Merging of observed states into clusters or crossing of hidden paths}
While lumping of microstates as well as post-milestoning lumped dynamics leads to some particular, counter-intuitive phenomena \cite{schwarzConsistentTimeReversal2025,blomMilestoningEstimatorsDissipation2024a}, we here discuss  the reverse process of post-lumping milestoned dynamics.
This merging of milestones into clusters of states (i.e.\ lumping of the milestones) generally breaks the semi-Markovian (renewal) dynamics.
It has been shown that certain models allow merging observed states into clusters 
to yield what is called a semi-Markov process of second order \cite{mart19a,Hartich2023PRR,blomMilestoningEstimatorsDissipation2024a}
and even higher $k$th order \cite{schwarzConsistentTimeReversal2025,zhaoMarkovstateHolography2025}, where the splitting probability for the next mesostate depends on the preceding $k-1$ states. By generalizing the notion of (semi)Markov order in the sense of convergence to a notion of \emph{weak (semi)Markov order}, one can even describe (semi)Markov processes of (strongly) infinite order emerging from a coarse-graining of mixing Markov processes \cite{zhaoMarkovstateHolography2025}. While some notable advances have been made in understanding and estimating the affinity contribution to $\dS$ from such dynamics \cite{schwarzConsistentTimeReversal2025}, much less is known about the underlying waiting time-statistics (see, however, works focusing on waiting times  between ``tagged'' transitions, e.g., \cite{vandermeerThermodynamicInferencePartially2022a,fritzEntropyEstimationPartially2025}.)

\subsection{Examples}

In this section we provide some illustrative examples on how to use and apply the developed theory.

\subsubsection{Hidden cycle in non-equilibrium system} \label{sec:3-state-example}

\begin{figure}

\begin{tikzpicture}

\pgfdeclaredecoration{sl}{initial}{
  \state{initial}[width=\pgfdecoratedpathlength-1sp]{
     \pgfmoveto{\pgfpointorigin}
  }
  \state{final}{
     \pgflineto{\pgfpointorigin}
    }
}

\tikzset{parallel arrow/.style={-latex,thick, 
     shorten >=1.5mm, shorten <=1.5mm, 
     decoration={sl,raise=1mm},decorate}}

\tikzset{thick parallel arrow/.style={-latex,very thick, 
     shorten >=2mm, shorten <=2mm, 
     decoration={sl,raise=1mm},decorate}}

	 \node[draw,thick,circle,fill=blind2,minimum size =0.6cm] at  (0,0) (1) {};
	 \node[draw,thick,circle,fill=blind2,minimum size =0.6cm] at  (10,0) (2) {};
	 
	  \node[draw,thick, dotted,circle,minimum size =0.6cm] at  (5,-4) (a) {};

	 \path[black!50] (1) edge[thick parallel arrow] node[above=0.5mm,black]{$\exp(E_1)$} (2);
	 \path[black!50] (2) edge[thick parallel arrow] node[below=0.5mm,black]{$\exp(E_2)$} (1);

	 \begin{scope}[black!50]
	 \path (1) edge[parallel arrow] node[above=0.5mm,sloped,black]{$\lambda \exp(\mu+E_1)$} (a);
	 \path (a) edge[parallel arrow] node[below=0.5mm,sloped,black]{$\lambda \exp(B)$} (1);
	 \path (2) edge[parallel arrow] node[above=0.5mm,sloped,black]{$\lambda \exp(B)$} (a);
	 \path (a) edge[parallel arrow] node[below=0.5mm,sloped,black]{$\lambda \exp(E_2)$} (2);

	 \draw[blind1,very thick, dashed, rounded corners =4.5mm] 
	 ($(a.north)+(-4,3)$)--($(a.north)+(+4,+3)$)--($(a.south)+(+4,-0.5)$)--($(a.south)+(-4,-0.5)$)--cycle;
	 \node[blind1,font=\large] at ($(a.south)+(+2,0)$) {$\Omega_{\rm hidden}^{1,2}$};
	 \end{scope}

	\node at (1) {\large $1$};
	\node at (2) {\large $2$}; 
	
	\node at (a) {\large $3$};

    \node[font=\large,anchor=south] at (0,0.5) {observed state 1};
    \node[font=\large,anchor=south] at (10,0.5) {observed state 2};
    \node[blind1,font=\large,anchor=north] at (5,-5) {unobserved state (chemically modified)};

\end{tikzpicture}
 \caption{The unobservable non-equilibrium cycle. Transitions can happen either directly from state $1$ to $2$, or via the hidden state $3$, so that only transitions between two states and thereby no cyclic transitions are observable. From the transition-path time distributions, it is possible to infer model parameters and thereby the driving of the unobservable cycle. We provide a detailed analysis thereof in Section~\ref{sec:3-state-example}
 } 
 \label{fig:3-state-example}
\end{figure}
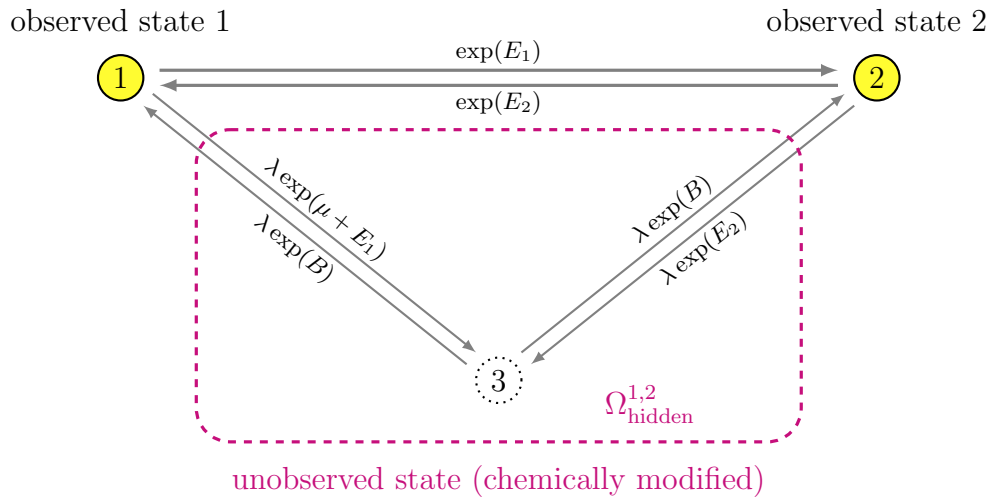

We consider the following 3-state non-equilibrium system with a hidden cycle. In particular, we have $\Omega_{\rm obs}= \{1,2\}$ and $\Omega_{\rm hidden}=\{3\}$, where we have a direct transition between states $1$ and $2$ as well as an indirect transition via state $3$. 
The transition $1 \rightarrow 3$ is driven by parameter $\mu$, as depicted in Fig.~\ref{fig:3-state-example}. Importantly, due to the milestoning, this driven cycle is not observable. This is a minimal example where both detailed balance violation and a hidden mechanism occur.

We thus have the following direct jumps [transitions of type (iii)] with out-rates
\begin{align}
	\begin{aligned}
w_{1 \rightarrow 2} 
&= \exp(E_1),
& \qquad w_{2 \rightarrow 1} &= \exp(E_2)\\
r^{(1)} 
&= \exp(E_1),
& r^{(2)} &= \exp(E_2).
\end{aligned}
\end{align}
Since $|\Omega_{\rm hidden}|=1$, we have no type-(i) jumps and the generator of the hidden dynamics is $\mathbf L^{1, 2}=0$. The type-(ii) jumps are 
\begin{align}
	\begin{aligned}
	b_3^{(1 \mid 2)} 
	&= \lambda \exp(B), 
	&  \qquad
	b_3^{(2 \mid 1)} 
	&=	\lambda \exp(B)
	\\
	c_3^{(1 \mid 2)} 
	&= \lambda \exp(E_2), 
	& 
	c_3^{(2 \mid 1)} 
	&=	\lambda \exp(E_1 +\mu).
	\end{aligned}
\end{align} 
Following Eq.~\eqref{eq:splitting_dec}, we can determine the splitting probabilities
\begin{align}
	\phi_{2|1}^{\rm jump} &= \frac{w_{1 \rightarrow 2}}{r_1+ b_{3}^{(1 \mid 2)} \left( b_{3}^{(1 \mid 2)} + b_{3}^{(1 \mid 2)} - L^{1,2} \right)^{-1} c_3^{(2 \mid 1)}} = \frac{2}{2+\lambda \exp(\mu)}. \label{eq:ex:hidden-cycle:phi-jump-2-given-1}
\end{align}
Analogously, we obtain
\begin{align}
	\begin{aligned}
	\phi_{2|1}^{\rm hidden} 
	&= \frac{\lambda \exp(\mu)}{2+\lambda \exp(\mu)},
	& \qquad
	\phi_{1|2}^{\rm jump} &= \frac{2}{2+\lambda},
	& \qquad
	\phi_{1|2}^{\rm hidden} &= \frac{\lambda}{2+\lambda},
	\end{aligned}  \label{eq:ex:hidden-cycle:phi-jump-rest}
\end{align}
which is consistent ($\phi_{2|1}^{\rm hidden} + \phi_{2|1}^{\rm jump} = \phi_{1|2}^{\rm hidden} + \phi_{1|2}^{\rm jump} = 1$).

\begin{figure}
 \includegraphics[width=\textwidth]{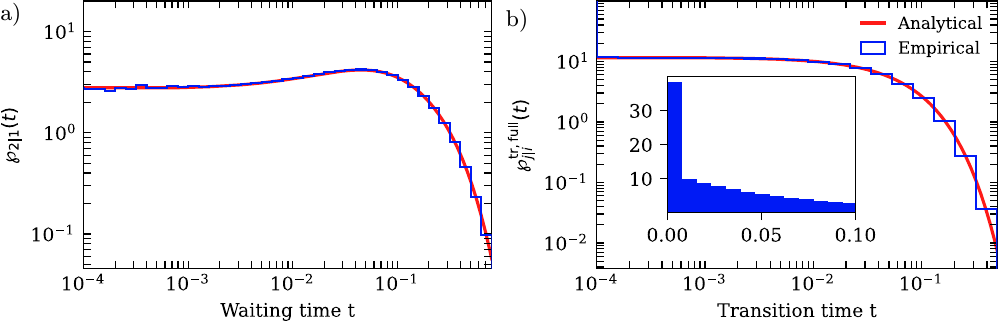}
  \caption{Density of (a) the waiting time and (b) the transition time in the unobserved non-equilibrium cycle of transitioning from observed state $1$ to observed state $2$: The analytical solution (red line) matches the measured waiting time distribution (blue). The direct jump $\phi_{2|1}^{\rm jump}\delta(t)$ can be seen at time $t=0$ in the inset of (b). We considered a trajectory of $10^7$ steps, after simulating initial $10^7$ steps for equilibration. Parameters are $E_1 =1$, $E_2 = 3$, $B=5$, $\lambda = \exp(-3)$, and $\mu = 5$.}
 \label{fig:3-state-system:waiting-time}
\end{figure}

The local generator from state $1$ is on the state space $\Omega^{(1)}=\{1,3\}$ and reads
\begin{equation}
 \mL^{(1)}=
 \begin{pmatrix}
 -\lambda \exp(E_1+\mu) & \lambda \exp(B) \\
 \lambda \exp(E_1 + \mu) & -\lambda \exp(B)
 \end{pmatrix}-
  \begin{pmatrix}
  \exp(E_1) & 0 \\
  0 & \lambda \exp(B) 
 \end{pmatrix}.
\end{equation}
By Eq.~\eqref{eq:wp} we obtain the \emph{waiting time density}
\begin{equation}
 \wp_{2|1}(t)=
\left(
\exp(E_1),\lambda \exp(B)\right)
\exp(\mL^{(1)}t)
 \begin{pmatrix}
  1\\0
 \end{pmatrix}
,
 \end{equation}
 which is quantitatively corroborated by numerical experiments shown in Fig.~\ref{fig:3-state-system:waiting-time}a.

The \emph{transition} time has density
\begin{equation}
	 \wp^{\rm {tr, full}}_{j|i}(t) = \phi_{j|i}^{\rm jump} \delta(t) + \phi_{j|i}^{\rm hidden}2 \lambda \exp(B)  \exp(-2 \lambda \exp(B) t),
\end{equation} 
which agrees with Eq.~\eqref{eq:wptrans} since we obtain from Eq.~\eqref{eq:ptrans} 
\begin{align}
\begin{aligned}
  \psi^{\rm tr}_{2|1}(t)
  &=\frac{(\bbb^{(1|2)})^\T\big[\exp\left((\mL^{1,2}-\mB^{(2|1)}-\mB^{(1|2)})t\right)\big]\bc^{(2|1)}}{(\bbb^{(1|2)})^\T\big[\mB^{(2|1)}+\mB^{(1|2)}-\mL^{1,2}\big]^{-1}\bc^{(2|1)}}\\
  &=2 \lambda \exp(B) \exp(-2 \lambda \exp(B) t).
  \end{aligned}
\end{align}
Because we have both, hidden mechanisms and driving, we actually have $ \wp^{\rm {tr, full}}_{2|1}(t) \neq  \wp^{\rm {tr, full}}_{1|2}(t)$. In particular, this stems from $ \phi_{2|1}^{\rm jump} \neq  \phi_{1|2}^{\rm jump}$ and $\phi_{2|1}^{\rm hidden} \neq \phi_{1|2}^{\rm hidden}$.
Moreover, numerically, we can see in Fig.~\ref{fig:3-state-system:waiting-time}b the jump at $t=0$ resulting from  $\phi_{2|1}^{\rm jump}\delta(t)$. 

Using Eqs.~\eqref{eq:ex:hidden-cycle:phi-jump-2-given-1}-\eqref{eq:ex:hidden-cycle:phi-jump-rest} we find that the chemical driving is given by
\begin{align}
	\mu = \ln \frac
	{\phi_{2|1}^{\rm hidden}  \phi_{1|2}^{\rm jump} }
	{\phi_{1|2}^{\rm hidden} \phi_{2|1}^{\rm jump} }.
	\label{eq:3-system:splittings-and-mu}
\end{align}

\begin{figure}
 \centering
 \includegraphics{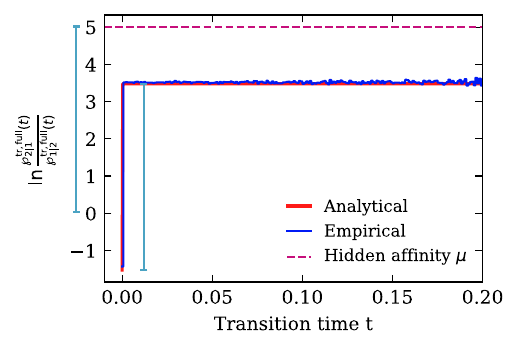}
   \caption{Log ratio of forward versus backward transition densities $\wp^{\rm {tr, full}}_{j|i}(t)$ (analytical results in red, binned simulation results in blue) for the system depicted in Fig.~\ref{fig:3-state-example}. From the values at $t=0$ and at any other $t>0$, we can exactly infer the driving $\mu = 5$ (this corresponds to the maximal difference of the ratio, teal bars). We considered a trajectory of $10^8$ steps, after simulating initial $10^7$ steps for equilibration, and used bin size $\delta=10^{-3}$. For other parameters, see Fig.~\ref{fig:3-state-system:waiting-time}.}
 \label{fig:3-state-system:abs-log-ratios}
\end{figure}

Hence, the non-equilibrium driving $\mu$ can be deduced solely from the transition time density
\begin{align}
	\mu  =\lim_{\varepsilon \rightarrow 0} \left\{  \ln\left[\frac{\int_\epsilon^\infty \wp^{\rm {tr, full}}_{2|1}(t ) \d t }{\int_\epsilon^\infty \wp^{\rm {tr, full}}_{1|2}(t) \d t}\right]-
    \ln\left[\frac{\int_0^\epsilon \wp^{\rm {tr, full}}_{2|1}(t)\d t}{\int_0^\epsilon\wp^{\rm {tr, full}}_{1|2}(t)\d t}\right] \right\}.
	\label{eq:3-system:log-ratio-densities}
\end{align}
Note that this limit also corresponds to the maximal difference between the two log ratios, see Fig.~\ref{fig:3-state-system:abs-log-ratios}, where we show the empirical and theoretical log ratios and how they allow to infer the hidden affinity $\mu$. It can be seen in Fig.~\ref{fig:3-state-system:abs-log-ratios} that the (log) ratio of the densities is constant for every $t > 0$. This phenomenon can also be derived analytically: From the 
reflection symmetry of transition-path times $\psi^{\rm tr}_{2|1}(t) = \psi^{\rm tr}_{1|2}(t)$ (see Section~\ref{sec:symm-transition-path-times}) and the definition of $\wp^{\rm {tr, full}}_{j|i}(t)$ (Eq.~\eqref{eq:wptrans}), we see that the ratio of densities is constant for any $t>0$. For parameter inference, this insight can be used to simplify the improper integral in Eq.~\eqref{eq:3-system:log-ratio-densities} to a simple ratio over bins, i.e.\ for any $\tau>0$ we have with bin width $\delta$
\begin{align}
    \lim_{\varepsilon \rightarrow 0} \ln\left[\frac{\int_\epsilon^\infty \wp^{\rm {tr, full}}_{2|1}(t ) \d t }{\int_\epsilon^\infty \wp^{\rm {tr, full}}_{1|2}(t) \d t}\right] = 
    \lim_{\delta \rightarrow 0} 
    \ln\left[\frac{\int_{\tau-\delta/2}^{\tau+\delta/2} \wp^{\rm {tr, full}}_{2|1}(t) \d t}{\int_{\tau-\delta/2}^{\tau+\delta/2} \wp^{\rm {tr, full}}_{1|2}(t) \d t}\right].
\end{align}

In the system considered in Fig.~\ref{fig:3-state-example} we can infer $\mu$ exactly. In experiments on more complex systems, say with cycles on multiple different time scales, one may furthermore use a recursive approach to infer the different affinities at different time scales, which will be left for future work.

\subsubsection{Two hidden paths}\label{sec:two-hidden-paths}

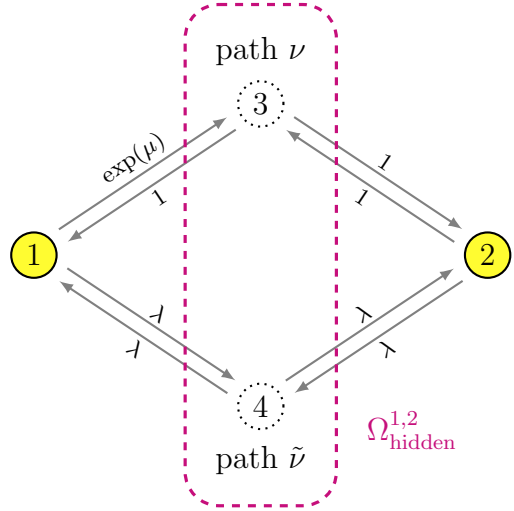
\begin{figure}

\begin{tikzpicture}

\pgfdeclaredecoration{sl}{initial}{
  \state{initial}[width=\pgfdecoratedpathlength-1sp]{
     \pgfmoveto{\pgfpointorigin}
  }
  \state{final}{
     \pgflineto{\pgfpointorigin}
    }
}

\tikzset{parallel arrow/.style={-latex,thick, 
     shorten >=1.5mm, shorten <=1.5mm, 
     decoration={sl,raise=1mm},decorate}}

\tikzset{thick parallel arrow/.style={-latex,very thick, 
     shorten >=2mm, shorten <=2mm, 
     decoration={sl,raise=1mm},decorate}}

 \node[font=\large,anchor=south] at (3,2.4) {path $\nu$};
  \node[font=\large,anchor=north] at (3,-2.4) {path $\tilde \nu$};
     
	 \node[draw,thick,circle,fill=blind2,minimum size =0.6cm] at  (0,0) (1) {};
	 \node[draw,thick,circle,fill=blind2,minimum size =0.6cm] at  (6,0) (2) {};

	  \node[draw,thick, dotted,circle,minimum size =0.6cm] at  (3,+2) (a) {};
	  
	  \node[draw,thick, dotted,circle,minimum size =0.6cm] at  (3,-2) (b) {};

	 \begin{scope}[black!50]
	 \path (1) edge[parallel arrow] node[above=0.5mm,sloped,black]{$\exp(\mu)$} (a);
	 \path (a) edge[parallel arrow] node[below=0.5mm,sloped,black]{$1$} (1);
	 \path (2) edge[parallel arrow] node[above=0.5mm,sloped,black]{$1$} (a);
	 \path (a) edge[parallel arrow] node[below=0.5mm,sloped,black]{$1$} (2);
	 
	 \path (1) edge[parallel arrow] node[above=0.5mm,sloped,black]{$\lambda$} (b);
	 \path (b) edge[parallel arrow] node[below=0.5mm,sloped,black]{$\lambda$} (1);
	 \path (2) edge[parallel arrow] node[above=0.5mm,sloped,black]{$\lambda$} (b);
	 \path (b) edge[parallel arrow] node[below=0.5mm,sloped,black]{$\lambda$} (2);

	 \draw[blind1,very thick, dashed, rounded corners =4.5mm] 
	 ($(a.north)+(-1,+1.0)$)--($(a.north)+(+1,+1.0)$)--($(b.south)+(+1,-1.0)$)--($(b.south)+(-1,-1.0)$)--cycle;
	 \node[blind1,font=\large] at ($(b.south)+(+2,0)$) {$\Omega_{\rm hidden}^{1,2}$};

	 \end{scope}

	\node at (1) {\large $1$};
	\node at (2) {\large $2$}; 
	
	\node at (a) {\large $3$};
	\node at (b) {\large $4$};
\end{tikzpicture}
 \caption{The system with two hidden transition mechanisms: Path $\nu$ connects state $1$ to state $2$ via hidden state $3$. Path $\tilde \nu$ connects via hidden state $4$. We provide a detailed analysis of this system in Section~\ref{sec:two-hidden-paths}.}
 
 \label{fig:4-state-example}
\end{figure}

We consider the 4-state-system with two hidden states $\Omega_{\rm hidden}=\{3,4\}$ depicted in Fig.~\ref{fig:4-state-example}. That is, we have only type-$(ii)$-transitions, but now two different parallel mechanisms to transition between the two observable states $\Omega_{\rm obs}=\{1,2\}$. We call the transition via state $3$ mechanism $\nu$, while the transition via $4$ is $\tilde \nu$. Since $|\Omega_{\rm hidden}^{1,2}(\nu)|= |\{3\}| = 1$, the generator $\mL^{1,2}(\nu) = 0$ is simply the scalar $0$, matching our observation of no type-$(i)$-transitions. Analogously, $\mL^{1,2}(\tilde \nu) = 0$.

For the type-$(ii)$ transitions, we get along path $\nu$
\begin{align}
	\begin{aligned}
	b_3^{(1 \mid 2)} 
	&= w_{3 \rightarrow 2} = 1, 
	& \qquad
	b_3^{(2 \mid 1)} 
	&=	w_{3 \rightarrow 1} = 1
	\\
	c_3^{(1 \mid 2)} 
	&=w_{2 \rightarrow 3} = 1, 
	&
	c_3^{(2 \mid 1)} 
	&=w_{1\rightarrow 3} = \exp(\mu),
	\end{aligned}
\end{align} 
and along path $\tilde \nu$

\begin{align}
	\begin{aligned}
	b_4^{(1 \mid 2)} 
	&= w_{4 \rightarrow 2} = \lambda, 
	& \qquad
	b_4^{(2 \mid 1)} 
	&=	w_{4 \rightarrow 1} = \lambda
	\\
	c_4^{(1 \mid 2)} 
	&=w_{2 \rightarrow 4} = \lambda, 
	&
	c_4^{(2 \mid 1)} 
	&=w_{1 \rightarrow 4} = \lambda.
	\end{aligned}
\end{align} 

Since there are no type-$(iii)$-transitions, we have $r^{(1)}=r^{(2)}=0$. 

\begin{figure}
 \includegraphics[width=\textwidth]{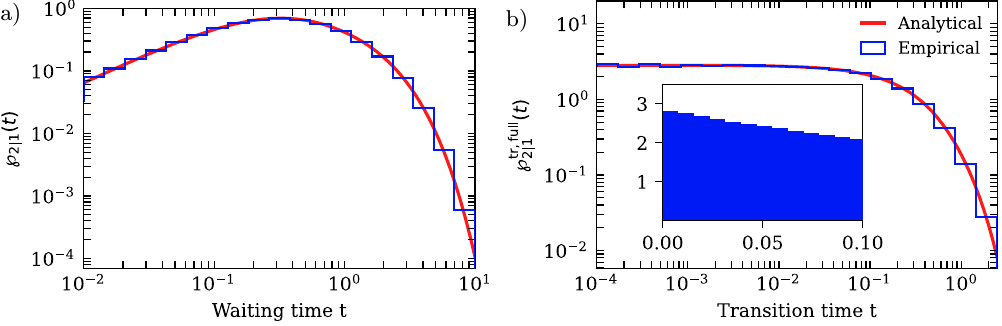}
  \caption{Density of (a) the waiting time and (b) the transition time in the 4-state-model (Fig.~\ref{fig:4-state-example})
  of transitioning from state $1$ to state $2$: The analytical solution (red) matches the measured waiting time and transition time density (blue). In contrast to the 3-state-system, we observe no direct jumps at $t=0$ (inset of (b)), which is consistent with our theory. We considered a trajectory of $10^8$ steps, after simulating initial $10^7$ steps for equilibration. Parameters are $\mu = 1$ and $\lambda = 2$.}
 \label{fig:4-state-system:waiting-time}
\end{figure}

The local generator from state $1$ is on the state space $\Omega^{(1)}=\{1,3,4\}$ and reads
\begin{equation}
 \mL^{(1)}=
 \begin{pmatrix}
 -\exp(\mu)-\lambda  &  1  & \lambda \\
 \exp(\mu)     & -1  &  0\\
 \lambda & 0 & -\lambda
 \end{pmatrix} 
 -  
 \begin{pmatrix}
 0  &  0  &  0\\
 0  &  1  &  0\\
 0  &  0  &  \lambda 	
 \end{pmatrix}
.
\end{equation}
Generalizing Eq.~\eqref{eq:wp} to multiple parallel mechanisms, we obtain for the \emph{waiting time density}
\begin{equation}
 \wp_{2|1}(t)=
\left(
0,1, \lambda \right)
\exp(\mL^{(1)}t)
 \begin{pmatrix}
  1\\0\\0
 \end{pmatrix}
,
 \end{equation}
 where the first row-vector stems from the generalization
 \begin{align}
 	\left(w_{1 \rightarrow 2}, \left[\bbb^{(1|2)}(\nu)\right]^\T,\left[\bbb^{(1|2)}(\tilde \nu)\right]^\T  \right).
 \end{align}
 The density is fully corroborated by the numerical experiments, as shown in Fig.~\ref{fig:4-state-system:waiting-time}a.

For the \emph{transition} probability distribution, we first note that due to the absence of direct transitions, we have $\phi_{2|1}^{\rm jump} = 0$ and $\phi_{2|1}^{\rm hidden} = 1$.
However, we now have two parallel mechanisms. We therefore consider both paths separately. 

Along path $\nu$, we have by Eq.~\eqref{eq:ptrans}
\begin{align}
	\psi^{\rm tr}_{2|1}(t,\nu) = 2 \exp(-2t)
\end{align}
The probability of having gone along $\nu$ given that we transitioned from state $1$ to state $2$ is the total mass of the waiting time density given that we transition along $\nu$, that is
\begin{align}
	\phi_{2 \mid 1}^{\rm hidden}(\nu) = \int_0^{\infty} \wp_{2|1}(t,\nu)\d t = \int_0^\infty \left(
0,1, 0 \right)
\exp(\mL^{(1)}t)
 \begin{pmatrix}
  1\\0\\0
 \end{pmatrix}\d t. \label{eq:ex-int-to-be-contd}
\end{align}
To evaluate this integral, we note that the local generator  $\mL^{(1)}$ may well be extended to a generator with a fourth state added as absorbing state
\begin{align}
    \mL^{(1)}_\text{absorb}=
 \begin{pmatrix}
 -\exp(\mu)-\lambda  &  1  & \lambda & 0 \\
 \exp(\mu)     & -2  &  0 & 0\\
 \lambda & 0 & -2\lambda & 0 \\
 0 & 1 & \lambda & 0
 \end{pmatrix}.
\end{align}
Since all states are connected, the Markov-Jump dynamics (i.e.\ $\lambda > 0$) is eventually  absorbed into the added fourth state, so that we have for any initial distribution vector $p$ that 
\begin{align}
    \lim_{t\rightarrow \infty} \exp(\mL^{(1)}_\text{absorb}t) \boldsymbol{p} = \boldsymbol{e_4},
\end{align}
where the convergence is exponential i.e.\ there exist $C > 0$ and the largest non-zero eigenvalue of $\mL^{(1)}_\text{absorb}$, $\lambda < 0$, such that for $\boldsymbol{e_2} \perp \boldsymbol{e_4}$ we have 
$$\boldsymbol{e_2}^\T  \exp(\mL^{(1)}_\text{absorb}t) \boldsymbol{p} = \boldsymbol{e_2}^\T \left( \exp(\mL^{(1)}_\text{absorb}t)\boldsymbol{p}-\boldsymbol{e_4}  \right) < C \exp(\lambda t) .$$
Hence, the following Laplace transform is indeed well defined.
\begin{align}
	f(s) 
    &\equiv \int_0^\infty \left(
0,1, 0 \right)
\exp(\mL^{(1)}t)
 \begin{pmatrix}
  1\\0\\0
 \end{pmatrix} \exp(-st)\d t  \\
 &= \left(
0,1, 0 \right)\int_0^\infty
\exp(\mL^{(1)}t)  \exp(-st)\d t
 \begin{pmatrix}
  1\\0\\0
 \end{pmatrix} \\
 &= \left(
0,1, 0 \right) \left(s\mI -  \mL^{(1)}\right)^{-1}\begin{pmatrix}
  1\\0\\0
 \end{pmatrix} 
\end{align}
and we now have
\begin{align}
    \lim_{s \rightarrow 0} \left(s\mI -  \mL^{(1)}\right)^{-1}
    = - (\mL^{(1)})^{-1}.
\end{align}
Since $\det( \mL^{(1)}) = -2 \lambda (\exp(\mu) + 2 \lambda) \neq 0$ for every  $\lambda>0$, the inverse is indeed well defined.
Hence, the conditions for $\int_0^\infty f(t)\d t=\lim_{s\to 0}\int_0^\infty {\rm e}^{-st}f(t)\d t$ are met and we obtain for Eq.~\eqref{eq:ex-int-to-be-contd}
\begin{align}
    	\phi_{2 \mid 1}^{\rm hidden}(\nu)  = \int_0^\infty \left(
0,1, 0 \right)
\exp(\mL^{(1)}t)
 \begin{pmatrix}
  1\\0\\0
 \end{pmatrix}\d t = \frac{\exp(\mu)}{\exp(\mu) + \lambda}.
\end{align}

Similarly, for $\tilde \nu$ we obtain  
\begin{align}
	\psi^{\rm tr}_{2|1}(t,\tilde \nu) = 2\lambda \exp(-2\lambda t),
\end{align}
and
\begin{align}
	\phi_{2 \mid 1}^{\rm hidden}(\tilde \nu) = \int_0^{\infty} \wp_{2|1}(t,\tilde \nu)\d t = \int_0^\infty \left(
0,0, \lambda \right)
\exp(\mL^{(1)}t)
 \begin{pmatrix}
  1\\0\\0
 \end{pmatrix}\d t = \frac{\lambda}{\exp(\mu) + \lambda},
\end{align}
where we note that the row vector now has a component $\lambda$, since mechanism $\tilde \nu$ enters target state $2$ with rate $\lambda$.
Thus, we have an overall transition probability density of 

\begin{align}
	\psi^{\rm tr}_{2|1}(t) 
	&= \phi_{2 \mid 1}^{\rm hidden}(\nu) \psi^{\rm tr}_{2|1}(t,\nu) + \phi_{2 \mid 1}^{\rm hidden}(\tilde \nu)\psi^{\rm tr}_{2|1}(t,\tilde \nu)\\
	&= \frac{2}{\exp(\mu) + \lambda} \left( \exp(\mu) \exp(-2t) + \lambda^2 \exp(-2\lambda t) \right).\label{eq:4-state-system:psi-2-given-1}
\end{align}
We compare this density with numerical results in Fig.~\ref{fig:4-state-system:waiting-time}b.

\begin{figure}
\centering
 \includegraphics{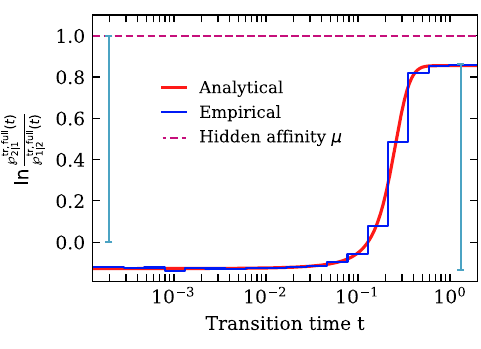}
  \caption{Inference of hidden cycle's affinity on multiple time scales: The microscopic cycle in the 4-state-model (Fig.~\ref{fig:4-state-example}) of affinity $\mu=1$ is hidden in the coarse-grained $2$-state-system. However, the depicted log ratio of transition times (analytic red line matches measured blue line)  allows to deduce the hidden affinity with and without time scale separation: The difference of the log ratio between small times and large times allows exact inference (teal bars). Measurements were done on trajectory of $10^8$ steps, after simulating initial $10^7$ steps for equilibration, with parameters $\mu=1$, $\lambda=10$.}
 \label{fig:4-state-system:inference-of-hidden-affinity}
\end{figure}

The reverse transition density $\psi^{\rm tr}_{1|2}(t) $ is obtained analogously to Eq.~\eqref{eq:4-state-system:psi-2-given-1}
\begin{align}
	\psi^{\rm tr}_{1|2}(t) 
	&= \frac{2}{\lambda+1} \left( \exp(-2t) + \lambda^2 \exp(-2\lambda t) \right).\label{eq:4-state-system:psi-1-given-2}
\end{align}
Similarly to Section \ref{sec:3-state-example}, we can infer the affinity $\mu$ as depicted in Fig.~\ref{fig:4-state-system:inference-of-hidden-affinity}.

\subsubsection{Robustness of milestone positioning and diffusion limit}\label{sec:robustness}

We now consider a particle diffusing in a potential on a ring under the effect of constant driving $M$. That is, the particle evolves on the domain $X \in [0,3)$ according to the Langevin equation 
\begin{align}
    \d X_t = f(x)\d t + \sqrt{2} \d W_t
\end{align}

with periodic boundary conditions, with
effective ``potential'' $U(x) = \frac{B}{2} (1-\cos(2 \pi x)) + M x$ with constants $B$ (barrier height) and M (driving intensity) and $f(x)\equiv -\partial_x U(x)$ on $x \in [0,3)$ such that $U(x)$ has a weak discontinuity at $x=3$. Without loss of generality we have set the diffusion constant $D \equiv 1$ and $k_{\rm B}T=1$.

\begin{figure}
\centering
 \includegraphics{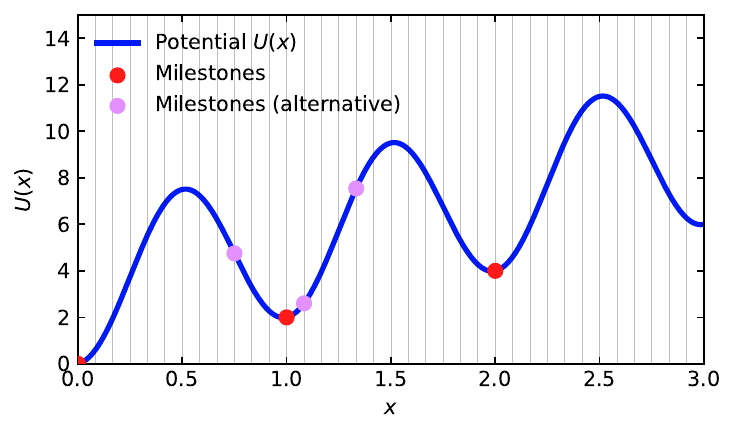}
  \caption{Effective periodic potential $U(x)$ is discretized to $36$ states (gray lines) with $3$ milestones (red and pink). The milestoning-based estimator of $\dS$ recovers the exact microscopic $\dS$ with 
  $\ge 3$ milestones, whose placement is arbitrary as the example contains a single dissipative cycle. That is, the result holds independently of the particular choice of milestones (as long as there are at least 3) whereby, remarkably, \emph{all 3 milestones} in the alternative milestoning (pink) are placed in the same potential well.  The driving parameter was set to $M \equiv 3 \pi$ and the barrier height was taken to be $B \equiv 6.5$.}
 \label{fig:discrete:potential}
\end{figure}

We may approximate the SDE through a continuous-time Markov chain with rates \cite{holu19}
\begin{align}
    r_{x \rightarrow (x\pm \d x \;\mathrm{mod}\; 3)}:= \frac{1}{\d x^2} \exp\left(-\frac{U(x\pm \d x) - U(x)}{2}\right),
    \label{eq:ex-disc-cont-limit:disc-rates}
\end{align}
which can be easily verified \cite{holu19,stutzerStochasticCalculusPathwise2025} (e.g.\ via the generator) to converge to the SDE. We depict a potential and its (milestoned) discretization in Fig.~\ref{fig:discrete:potential}.

We note that the eigenvalues of the generator of the periodic process interlace with the eigenvalues of the generator with one side being absorbing. Similarly, the eigenvalues of the generator with one state being absorbing interlace with the eigenvalues of the generator with both states being absorbing \cite{hart19}.

\begin{figure}[h!!]
\centering
 \includegraphics{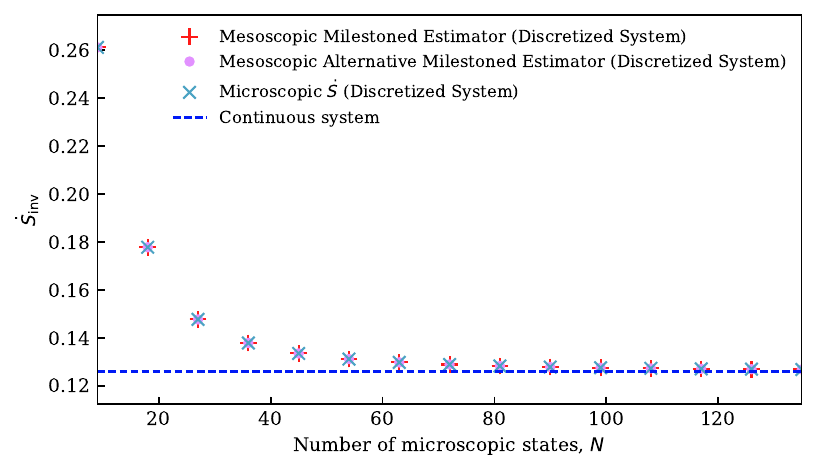}
  \caption{Analytical results for the entropy production rate of the discretized system as a function of the number of states $N$. The milestoning estimator (Eq.~\eqref{eq:entropy_preservation}) recovers the full entropy production rate of the discretized model already with only three states (milestones). The three-state milestoning estimate and the ``microscopic'' Markov estimate of the discretized Markov-state model (Eq.~\eqref{eq:entropy_def}) converge to the limiting value of $\dS$ of the continuous system (Eq.~\eqref{eq:ex-SDE:epr}) as $N\to\infty$ (dashed blue line). The particular choice of milestoning is irrelevant as long as there are at least 3, i.e., placing one milestone per well (red) or all milestones in the same well (pink) yields the same (exact) result as the ``microscopic'' system (teal).}
 \label{fig:discrete:convgs}
\end{figure}

We now consider the steady-state entropy production rate of the system. The 
corresponding Fokker-Planck equation with dimensionless drift $f(x)=-B \pi \sin(2\pi x) - M$ reads $\partial_tp(x,t)=-\partial_x J(x,t)$ with $J(x,t)=(f(x)-\partial_x)p(x,t)$ under periodic boundary conditions $p(x+3,t)=p(x,t)$. In the steady state we have $\partial_xJ=0$, i.e.\ $J^{\rm inv}={\rm const}$. A standard calculation (see Appendix~\ref{SSEPR}) yields
\begin{align}
    p^{\rm inv}(x)
    &= N^{-1}\exp(-U(x)) \left[ \int_x^3 \exp(U(y))\d y + \exp(3M)  \int_0^x \exp(U(y))\d y \right]\\
   J^{\rm inv}& = N^{-1}[1-\exp(3M)]
\end{align}
with normalization constant
\begin{align}
N=\int_0^3\exp(-U(x))\left[\int_x^3\exp(U(y))dy+\exp(3M)\int_0^x\exp(U(y))\d y \right]\d x.    
\end{align}
The steady-state entropy production rate thus reads
\begin{align}
    \dS^{\rm inv } = (J^{\rm inv})^2 \int_0^3  \frac{1}{p^{\rm inv}(x)} \d x. \label{eq:ex-SDE:epr}
\end{align}
With the probability current obtained in Eq.~\eqref{eq:ex-SDE:J-inv-through-C}, we determine from  \eqref{eq:ex-SDE:epr} the value of $\dS^{\rm inv }$
for $M\equiv 3 \pi$ and $B\equiv 6.5$, and compare the result with that obtained for the discrete Markov approximation with rates given by Eq.~\eqref{eq:ex-disc-cont-limit:disc-rates}
and with the corresponding milestoned result in Eq.~\eqref{eq:entropy_preservation}. The results are shown in Fig.~\ref{fig:discrete:convgs}, confirming the robustness to milestone positioning as well as their convergence to the continuous limit.

\section{Conclusion}
We derived an exact coarse graining of generic ergodic Markov-jump processes into 
semi-Markov (renewal) dynamics on an observable state space. We obtained exact results for 
waiting-time distributions for jumps between observable states and proved that these decompose into conditionally independent dwell and transition-path time periods. We furthermore proved that dwell time is a local property of the mesostate, whereas transition-path times depend on both, the initial and final states, and were shown to trigger the phenomenon of kinetic hysteresis. Transition-path times were furthermore shown to obey a reflection symmetry as soon as the sub-network connecting two adjacent observable states satisfies detailed balance. 

These symmetries were shown to imply, for any renewal process emerging from a coarse-graining of thermodynamically consistent Markov dynamics, that unequal waiting-time distributions for a transition from one into any pair of adjacent mesostates necessarily require kinetic hysteresis. In turn, kinetic hysteresis emerges if there exist unequal waiting-time distributions for a transition from one into a pair of adjacent mesostates. Since kinetic hysteresis is independent of the (ir)reversibility of the underlying microscopic dynamics, waiting-time contributions to dissipation emerging from a naive Markovian definition of time-reversal (i.e., so-called ``waiting-time entropy production'') do \emph{not} encode dissipation.    
Consistently, when all transitions become instantaneous, a Markov process emerges and, if the microscopic dynamics are such that between any pair of observed states there are no multi-edge instantaneous transitions, the emerging Markov process is also thermodynamically consistent.

We fully characterized the stationary properties of the milestoned dynamics and
proved the thermodynamic consistency of the coarse-graining and the resulting (semi-Markov) renewal process in the sense that it preserves the microscopic steady-state entropy production rate as long as no dissipative cycles become hidden by the coarse graining. Importantly, the milestoning estimator of the microscopic entropy production rate is experimentally accessible. 
Remarkably, we have also shown that even in the limit of a time-scale separation, where the observed dynamics becomes to a good approximation Markovian, the effect of kinetic hysteresis on the dissipation persists. 
This highlights further that accounting for kinetic hysteresis in defining the physically consistent time-reversal operation for semi-Markov dynamics is crucial for ensuring thermodynamic consistency.

We furthermore illustrated 
how our framework under given conditions may be used to deduce \emph{exactly} the affinity of a hidden cycle solely from the forward and backward transition densities, and demonstrated the robustness of the theory and resulting estimator to milestone positioning.

While the presented theory is complete under the stated conditions for renewal dynamics emerging from milestoning Markov jump processes, many important questions about the effects of milestoning remain open.
In particular, further research is needed on milestoning of already coarse-grained processes, such as lumping subsequently followed by milestoning. For such a hierarchical coarse graining procedure, a general proof for thermodynamic consistency -- analogously to the present proof -- is still missing \cite{blomMilestoningEstimatorsDissipation2024a,dieballPerspectiveTimeIrreversibility2025}.
Moreover, an extension of the $k$-th order Markov theory \cite{schwarzConsistentTimeReversal2025} to $k$-th order \textit{semi}-Markov processes (i.e.\ including sequence-dependent waiting times beyond two states) remains elusive. Finally, as illustrated with a minimal example, the developed theory lays the foundation for inferring affinities of hidden dissipative cycles in more complex coarsely observed networks from distributions of transition-path times.

\newpage 
\section*{Acknowledgments} Financial support of the European Research Council (ERC) under the European Union’s Horizon Europe research and innovation program (Grant Agreement No. 101086182 to A-G.) as well as the German Research Foundation
(DFG) through the Heisenberg Program (Grant GO 2762/4-1 project number 519908342 to AG), the EPSRC Centre for Doctoral Training in Mathematics of Random Systems: Analysis, Modelling and Simulation (EPSRC Grant EP/S023925/1) and Rhodes Trust (to T.~S.) are gratefully acknowledged.

\section*{Data availability}
The data that support the findings of this study are available from the corresponding author upon request.

\section*{Author contributions}
\noindent \textbf{T.~S.}: Conceptualization, Investigation, Writing -- Original Draft.
\textbf{D.H.}: Conceptualization, Investigation, Writing -- Original Draft.
\textbf{A.G.}: Conceptualization, Investigation, Writing -- Original Draft, Supervision.

\section*{Competing Interests}
The authors declare no competing interests.

\appendix
\section{Moments of waiting, dwell and  transition time from Laplace transform}
\subsection{Moments of dwell time}
In Eq.~\eqref{eq:twp_dwell} we showed that the dwell time statistics satisfies $ \twp^{\rm dwell}_{j|i}(s)= \twp^{\rm dwell}_{i}(s)$. Denoting the dwell time by $\tau$ the Laplace transform of the dwell time reads $\twp^{\rm dwell}_{i}(s)=\avg{\exp(-s\tau)}^{\rm dwell}_i$, which implies the $n$th moment of the dwell time to be given by
\begin{equation}
  \avg{\tau^n}_{i}\equiv(-1)^n\partial_s^n\twp^{\rm dwell}_{i}(s)\big|_{s=0}.
\end{equation}
Using Eq.~\eqref{eq:twp_dwell} we use the first two moments of dwell time to obtain
\begin{align}
  \begin{aligned}
\avg{\tau}^{\rm dwell}_{i}&=
\frac{
1
+\sum_k(\bbb^{(k|i)})^\T\big[\mB^{(k|i)}+\mB^{(i|k)}-\mL^{k,i}\big]^{-2}\bc^{(k|i)}
}{
r^{(i)} +\sum_k(\bbb^{(i|k)})^\T\big[\mB^{(k|i)}+\mB^{(i|k)}-\mL^{k,i}\big]^{-1}\bc^{(k|i)}
}\\
\avg{\tau^2}^{\rm dwell}_{i}&=2(\avg{\tau}^{\rm dwell}_{i})^2
-2\frac{
\sum_k(\bbb^{(k|i)})^\T\big[\mB^{(k|i)}+\mB^{(i|k)}-\mL^{k,i}\big]^{-3}\bc^{(k|i)}
}{
r^{(i)} +\sum_k(\bbb^{(i|k)})^\T\big[\mB^{(k|i)}+\mB^{(i|k)}-\mL^{k,i}\big]^{-1}\bc^{(k|i)}
}
  \end{aligned}
  \label{eq:moments_dwell}
\end{align}
where we exploited the structure $\twp^d_i(s)=f(0)/f(s)$ 
which yields $\partial_s\twp^d_i(s)=-f(0)f'(s)/f(s)^2$ and thus $\avg{\tau}_{i}=f'(0)/f(0)$ as well as
$\avg{\tau^2}_{i}=2[f'(0)/f(0)]^2-f''(0)/f(0)$. We further used that the denominators of Eqs.~\eqref{eq:twp_pre} and \eqref{eq:twp_dwell} are identical\footnote{
This means $f(s)=s+r^{(i)} +c^{(i)}- \sum_k(\bbb^{(k|i)})^\T\big[\id s+\mB^{(k|i)}+\mB^{(i|k)}-\mL^{k,i}\big]^{-1}\bc^{(k|i)}$ and $f(0)=r^{(i)}+\sum_k(\bbb^{(i|k)})^\T\big[\mB^{(k|i)}+\mB^{(i|k)}-\mL^{k,i}\big]^{-1}\bc^{(k|i)}$} and differentiation of the inverse of a matrix yields $\partial_s\mA(s)^{-1}=-\mA^{-1}\mA'\mA^{-1}$.

\subsection{Moments of transition time}
Direct transitions between observed states are instantaneous while indirect transitions $i\to j$ are distributed according to Eq.~\eqref{eq:tptrans}. 
The $n$th moment of indirect transition time becomes
\begin{align}
 \avg{\delta t^n}_{j|i,\text{hidden}}^{\rm tr}=(-\partial_s)^n\tilde\psi^{\rm tr}(s)\big|_{s=0}=\frac{n!(\bbb^{(i|j)})^\T\big[\mB^{(j|i)}+\mB^{(i|j)}-\mL^{i,j}\big]^{-(n+1)}\bc^{(j|i)}}{(\bbb^{(i|j)})^\T\big[\mB^{(j|i)}+\mB^{(i|j)}-\mL^{i,j}\big]^{-1}\bc^{(j|i)}}.
 \label{eq:moments_trans_hidden}
\end{align}
where we differentiated Eq.~\eqref{eq:tptrans}. Eq.~\eqref{eq:moments_trans_hidden} only accounts for the hidden non-instantaneous transitions. 
Using Eq.~\eqref{eq:splitting_dec} we can also include the instantaneous transitions  such that the  $n$th  moment of transition time of a transition $i\to j$ becomes (here $n\ge 1$)
\begin{align}
   \avg{\delta t^n}_{j|i}^{\rm tr}&=\frac{\phi_{j|i}^{\rm hidden}\avg{\delta t^n}_{j|i,\text{hidden}}^{\rm tr}+\phi_{j|i}^{\rm jump}\avg{\delta t^n}_{j|i,\text{jump}}^{\rm tr}}{\phi_{j|i}}
   \nonumber\\
   &=
   \frac{n!(\bbb^{(i|j)})^\T\big[\mB^{(j|i)}+\mB^{(i|j)}-\mL^{i,j}\big]^{-(n+1)}\bc^{(j|i)}}{w_{i\to j}+(\bbb^{(i|j)})^\T\big[\mB^{(j|i)}+\mB^{(i|j)}-\mL^{i,j}\big]^{-1}\bc^{(j|i)}},
   \label{eq:moments_trans}
\end{align}
where we used $\avg{\delta t^n}_{j|i,\text{jump}}^{\rm tr}=0$ for $n\ge 1$.

\subsection{Moments of conditional waiting time}
The Laplace transform of the  joint distribution of  waiting time  $t$ and state change $j$ (from $i$) is given by Eq.~\eqref{eq:twp}. The conditional density reads $\twp_{j|i}(s)/\phi_{j|i}=\avg{\exp(-st)}_{j|i}$ such that
\begin{equation}
 \avg{t^k}_{j|i}=(-1)^k\partial_s^k\twp_{j|i}(s)/\phi_{j|i}\big|_{s=0}.
\end{equation}
One might be tempted to ``na\"ively'' determine the moments of conditional waiting time from Eq.~\eqref{eq:twp} and \eqref{eq:splitting}, which turned out to be an unnecessary calculation \cite{hartichEmergentMemoryKinetic2021a}.
Instead we exploit the independence between transition and dwell time
$t=  \delta t+\tau$
manifested in Eq.~\eqref{eq:independence}.
Hence, the  moments of the waiting time along a state change $i\to j$
read
\begin{equation}
\begin{aligned}
  \avg{t}_{j|i}&= \avg{\delta t}_{j|i}^{\rm tr}+\avg{\tau}_{i}^{\rm dwell} \\
  \avg{t^2}_{j|i}&= \avg{\delta t^2}_{j|i}^{\rm tr}+2\avg{\delta t}_{j|i}^{\rm tr}\avg{\tau}_{i}^{\rm dwell}+\avg{\tau^2}_{i}^{\rm dwell}.
  &
\end{aligned}
\end{equation}
with the moments of dwell and transition time given in  Eqs.~\eqref{eq:moments_dwell} and \eqref{eq:moments_trans}.

\section{Effective transition rates in ergodic system} \label{app:effecitve-rates-ergodic}
Since we work with an ergodic system, i.e., we have only one communicating class in the microscopic Markov dynamics, the law of large numbers implies for $t \rightarrow \infty$ that the time-averaged number of transitions between any pair of milestones per unit time concentrates tightly around the respective mean value. Hence, the invariant properties of the mesoscopic milestoned process, i.e., the invariant distribution, average exit times, and transition rates between the milestones in the stationary state, correspond to those of a replica continuous-time Markov process on the milestoned state space $\Omega_{\rm obs}$, which are straightforward to determine. This replica continuous-time Markov process is defined in terms of the effective transition rates described in Eq.~\eqref{eq:effective-rates}.

\section{Steady-state entropy production of diffusion driven on a ring}\label{SSEPR}
We require the invariant density $p^{\rm inv}(x)$ and current $J^{\rm inv}$, which follow from $J^{\rm inv } = \mathrm{const}$ and periodic boundary conditions. To obtain $p^{\rm inv}$, we solve
\begin{align}
    J^{\rm inv} 
    &= f(x) p^{\rm inv }(x) - \partial_x [p^{\rm inv}(x)],
\end{align}
with the integrating factor $\mu(x) = \exp(U(x))$ for $U(x) = -\int_0^x f(y) dy$ 
as 
\begin{align}
    \partial_x[\mu(x) p^{\rm inv}(x)]= - \mu(x)  J^{\rm inv}.
\end{align}
Hence for a constant $C$ we have,
\begin{align}
    p^{\rm inv}(x) &=\exp(-U(x)) \left( C - J^{\rm inv} \int_0^x \exp(U(y))\d y \right),
\end{align}
The above is the general solution for the line. Due to periodicity $\mod 3$, we must have $\forall x \in [0,3): \; p^{\rm inv}(x+3) =p^{\rm inv}(x)$. Further, we impose $\int_0^3 p^{\rm inv}(x) \d x = 1$.
A necessary condition following from the periodicity is $p^{\rm inv}(0) = p^{\rm inv}(3)$, hence
\begin{align}
    p^{\rm inv}(0) =  C = \exp(-3M) \left[C -  J^{\rm inv} \int_0^3 \exp(U(y))\d y \right] = p^{\rm inv}(3) 
\end{align}
where we used $U(3) = 3M$.
Thus
\begin{align}
    C = \frac{1}{1-\exp(3M)}  \left( J^{\rm inv}\int_0^3 \exp(U(y))\d y \right),
\end{align}
or equivalently
\begin{align}
    J^{\rm inv} =C  \frac{1-\exp(3M)}{\int_0^3 \exp(U(y))\d y } \label{eq:ex-SDE:J-inv-through-C}
\end{align}
such that
\begin{align}
    p^{\rm inv}(x)
    &= C \exp(-U(x)) \left[ 1 -  \frac{1-\exp(3M)}{\int_0^3 \exp(U(y))dy}  \int_0^x \exp(U(y))\d y \right]\label{eq:ex-SDE:p-inv-upto-C}
\end{align}
where we obtain $C$ from $\int_0^3  p^{\rm inv}(x) \d x = 1$.

\newpage 
\section*{References}
\bibliography{Biballes}

 \end{document}